\pdfoutput=1
\documentclass[twocolumn,prc,aps,showkeys,superscriptaddress,nofootinbib,floatfix]{revtex4}

\usepackage{amssymb}
\usepackage{amsmath}
\usepackage{graphicx} % Include figure files
\usepackage{dcolumn}  % Align table columns on decimal point
\usepackage{rotating} 
\usepackage{subfigure} 
\usepackage{titlesec}
\usepackage{comment}
\usepackage{breakurl}

\graphicspath{ {Figures/} }

\usepackage{float}    % To make bordet aroung figures
\floatstyle{boxed}

\newcommand*{\justifyheading}{\raggedright}
\titleformat{\chapter}[display]
  {\normalfont\huge\bfseries\justifyheading}{\chaptertitlename\ \thechapter}
  {20pt}{\Huge}
\titleformat{\section}
  {\normalfont\Large\bfseries\justifyheading}{\thesection}{1em}{}
\titleformat{\subsection}
  {\normalfont\large\bfseries\justifyheading}{\thesubsection}{1em}{}
\titleformat{\subsubsection}
  {\normalfont\large\justifyheading}{\thesubsubsection}{1em}{}
\usepackage[english]{babel}

\begin{document}

\pagestyle{empty}
\pagestyle{plain}
%\linenumbers

\title{A Conceptual Design Study of a Compact Photon Source (CPS) for Jefferson Lab}

\author{D.~Day}
\affiliation{University of Virginia, Charlottesville, Virginia 22904, USA}

\author{P.~Degtiarenko}
\affiliation{Thomas Jefferson National Accelerator Facility, Newport News, Virginia 23606, USA}

\author{S.~Dobbs}
\affiliation{Florida State University, Tallahassee, Florida 32306, USA}

\author{R.~Ent}
\affiliation{Thomas Jefferson National Accelerator Facility, Newport News, Virginia 23606, USA}

\author{D.J.~Hamilton}
\affiliation{University of Glasgow, Glasgow G12 8QQ, Scotland, United Kingdom}

\author{T.~Horn}
\email[Contact email:]{hornt@cua.edu}
\affiliation{Catholic University of America, Washington, D.C. 20064, USA}
\affiliation{Thomas Jefferson National Accelerator Facility, Newport News, Virginia 23606, USA}

\author{D.~Keller}
\affiliation{University of Virginia, Charlottesville, Virginia 22904, USA}

\author{C.~Keppel}
\affiliation{Thomas Jefferson National Accelerator Facility, Newport News, Virginia 23606, USA}

\author{G.~Niculescu}
\affiliation{James Madison University, Harrisonburg, Virginia 22807, USA}

\author{P.~Reid}
\affiliation{Saint Mary’s University, Halifax, Nova Scotia, Canada}

\author{I.~Strakovsky}
\affiliation{George Washington University, Washington, D.C. 20052, USA}

\author{B. Wojtsekhowski}
\affiliation{Thomas Jefferson National Accelerator Facility, Newport News, Virginia 23606, USA}

\author{J.~Zhang}
\affiliation{University of Virginia, Charlottesville, Virginia 22904, USA}

\newpage
\date{\today}

\begin{abstract}

  This document describes the technical design concept of a compact
  high intensity, multi-GeV photon source.  Capable of producing
  10$^{12}$ equivalent photons per second this novel device will
  provide unprecedented access to physics processes with very small
  scattering probabilities such as hard exclusive reactions on the
  nucleon. When combined with dynamic nuclear polarized targets, its
  deployment will result in a large gain in polarized experiment
  figure-of-merit compared to all previous measurements. 
Compared to a traditional bremsstrahlung photon source the proposed concept presents several advantages, most significantly in providing a full intensity in a small spot at the target and in taking advantage of the narrow angular spread associated with high energy bremsstrahlung compare to the wide angular distribution of the secondary radiation to minimize the operational prompt and activation radiation dose rates.

\end{abstract}


%============================================================

\maketitle

%\linenumbers

\section{Introduction}

A quantitative description of the nature of strongly bound systems is
of great importance for an improved understanding of the fundamental
structure and origin of matter.  One of the most promising ways to
access information on the dynamical structure of the nucleon is
through exclusive reactions at high momentum transfer, in which the
deep interior of the nucleon is probed with a highly-energetic photon
or electron probe and all final-state particles are
detected~\cite{Hand63,Kugler71}.  Even though the scattering
probability of such reactions is extremely small it has become clear
that such reactions offer a promising route to imaging of the elusive
3-D nucleon substructure.  Indeed, there have been increasingly
sophisticated theoretical efforts to exploit the richness of exclusive
reactions at short resolution scales~\cite{Goeke01}.

Exclusive measurements with high-energy electron and photon beams form
the core of the new paradigm within sub-atomic science termed "nuclear
femtography".  In both photon and electron scattering experiments, the
scale of the associated imaging that can be performed is set by the
invariant squared four-momentum transferred to the proton target,
$-t$, and the total centre-of-mass energy squared, $s$.  Measurements
over a wide range of $s$ and $-t$ with these probes allow for the
disentangling of four functions representing the vector, axial,
tensor, and pseudo-scalar response of the nucleon. Simultaneous
experimental access to all of these functions is most readily achieved
with a spin polarized nuclear or nucleon target.

Much progress imaging nucleon structure can be made with
electron-scattering reactions, yet experiments utilizing high-energy
photons play a unique complementary role. Measurements involving the
small scattering probabilities associated with exclusive reactions
demand high-intensity photon beams. Further, our basic understanding
will be much strengthened by imaging longitudinally-polarized and
transversely-polarized nucleons. It is for this combination that the
proposed concept is primarily focused: with a newly-developed compact
photon source (CPS) \cite{SABW2014, BWGN2015} and a dynamically-nuclear polarized target system,
for example in Hall C at Jefferson Lab,
a gain of a factor of 30 in the figure-of-merit (as defined by the
photon intensity and the average target polarization over the
experiment) can be achieved.  The net gain makes it possible to
measure the very small scattering cross sections associated with a new
suite of high-energy photon scattering experiments to image and
understand the dynamical nucleon structure~\cite{CPSWS}.

The concept of a CPS also enables other science possibilities, like
enriching the hadron spectroscopy program in Hall D at Jefferson Lab
and at other facilities.  Hall D is a newly-built experimental hall,
with a large acceptance spectrometer and a tagged, linearly polarized
photon beam of low to moderate intensity.  The addition of a CPS to
this hall opens the door to increased sensitivity to rare processes
through a higher intensity photon beam or the production of secondary
beams of other particles, such as a $K_L$ beam~\cite{Klong}.  Although
there are fewer physical limitations on the size of the CPS in Hall D,
allowing for additional flexibility in the optimization of the
shielding, most of the other requirements are similar to CPS running
in the other halls.  The radiation shielding requirements are similar
in order to ensure safe operation and to prevent radiation damage to
the tagger detectors and their associated electronics located upstream
of the planned CPS location.

For operation of the proposed $K_L$ facility, the electron beam has
been proposed to have a power up to 60kW, running at an energy of 12
GeV with a 64 ns beam bunch spacing.  Initial estimates suggest that
the default CPS configuration can handle the power deposition, and
sufficient cooling water is available, as the electron dump for the
nominal Hall D photon beam is designed to absorb at least 60 kW of
power.  A major difference,
as compared to Hall C,
is that the Hall D CPS is located in a
separate section of the hall from the target and main spectrometer,
and is separated by $\sim80$~m of pipe under vacuum surrounded by
soil.  The size of the photon beam generated by the CPS is dominated
by multiple scattering in the radiator, and has estimated to be 2~cm
after traveling 80~m.  This is well within the size of the
15~cm-diameter beam pipe, and the 6~cm-diameter Be $K_{L}$ target.
Finally, if the CPS radiator is retracted, then the current Hall D
photon beam can be used without moving the CPS or any other
modification from the beamline.  Taking all of these factors into
account, the CPS design is well matched for experiments in Hall D
requiring a high-intensity untagged photon beam.

\section{Science Opportunities with CPS}
\label{sec:science-gain}

Investigating the three-dimensional structure of the nucleon has
historically been an active and productive field of research,
especially so during the last two decades since the invention of the
generalized parton distributions (GPD) formalism. Research focused on
this three-dimensional structure continues to be central to the hadron
physics program at facilities like Jefferson Lab.  The GPD formalism
provides a unified description of many important reactions including
elastic electron scattering, deep-inelastic scattering (DIS),
deeply-virtual and timelike Compton scattering (DVCS and TCS),
deeply-virtual meson production (DVMP), and wide-angle real Compton
scattering (RCS) and meson production. All of these can be described
by a single set of four functions $H$, $\tilde{H}$, $E$ and
$\tilde{E}$, which need to be modeled and constrained with parameters
extracted from experimental
data~\cite{Diehl03,Burkardt02,Goeke01,Belitsky05,Ji97,Radyushkin96,Mueller94,Collins97,Collins99}.
The CPS science program as proposed for Jefferson Lab enables studies
of the three-dimensional structure of the nucleon and features one
fully approved and two conditionally approved
experiments~\cite{Klong,E12-17-008,C12-18-005}.

Jefferson Lab Experiment E12-17-008~\cite{E12-17-008} will measure
polarization observables in real Compton scattering (RCS).  This is a
fundamental and basic process, yet its mechanism in the center-of-mass
energy regime of $\sqrt{s}$ = 5-10 GeV remains poorly understood.
Measurements show that these data cannot be described by perturbative
calculations involving the scattering of three valence quarks. Rather
the dominant mechanism is the so-called "handbag model" where the
photon scatters from a single active quark and the coupling of this
struck quark to the spectator system is described by
GPDs~\cite{Rad98,Diehl99}. It is this latter conceptual mechanism that
lies at the root of the worldwide efforts of 3D (spatial) imaging of
the proton's quark-gluon substructure, as the GPDs contain information
about the transverse spatial distribution of quarks and their
longitudinal momenta inside the proton.

The RCS experimental observables provide several constraints for GPDs
which are complementary to other exclusive reactions due to an $e_a^2$
factor and an additional $1/x$ weighting in the corresponding GPD
integrals.  For example, the elastic form factor $F_1(t)$ is related
to the RCS vector form factor $R_V(t)$, both of which are based on the
same underlying GPD $H(x,0,t)$.  Similarly, polarized observables in

RCS uniquely provide high $-t$ constraints on $\tilde H(x,0,t)$ via
extraction of the RCS axial form factor $R_A(t)$ in a kinematic regime
where precise data on the nucleon axial form factor is not
available~\cite{Kroll17,Kroll18}.  A measurement of the spin asymmetry
in RCS with the proton target longitudinally polarized can further
disentangle the various reaction mechanism models. If consistent with
the measurement of the spin transfer from the photon to the scattered
proton, the asymmetry can be surprisingly large and stable with
respect to the photon center-of-mass scattering angle.  Investigations
into the mechanisms behind RCS will provide crucial insight into the
nature of exclusive reactions and proton structure and are ideally
suited for the facilities provided by the Jefferson Lab 12-GeV
upgrade~\cite{Dudek12,NPSWP14,DOE15,Mont17}.

Jefferson Lab Experiment C12-18-005~\cite{C12-18-005} will probe 3D
nucleon structure through timelike Compton scattering, where a real
photon is scattered off a quark in the proton and a high-mass
(virtual) photon is emitted, which then decays into a lepton
pair~\cite{Berger02,Anikin18}.  Using a transversely polarized proton
target and a circularly polarized photon beam allows access to several
independent observables, directly sensitive to the GPDs, and in
particular the $E$ GPD which is poorly constrained and of great
interest due to its relation to the orbital momentum of the
quarks~\cite{Kroll10,Kroll11,Gold12}. The experiment involves
measurements of the unpolarized scattering probabilities or cross
section, the cross section using a circularly polarized photon beam,
and the cross section using transversely-polarized protons. This will
provide a first fundamental test of the universality of the GPDs, as
the GPDs extracted from TCS should be comparable with those extracted
from the analogous spacelike (electron) scattering process -- deeply
virtual Compton scattering, a flagship program of the 12-GeV Jefferson
Lab upgrade~\cite{Dudek12,NPSWP14,DOE15,Mont17}.

A separate window on the nature of strongly bound systems is provided
through the hadron spectrum. The spectrum allows study of the properties
of QCD in its domain of strong-coupling, leading to the most striking feature
of QCD: the confinement of quarks and gluons within hadrons: mesons and baryons.
Experimental investigation of the baryon spectrum provides one obvious avenue to
understand QCD in this region since the location and properties of the excited states
depend on the confining interaction and the relevant degrees of freedom of hadrons.
 
Understanding the constituent degrees of freedom in hadrons requires
identifying a spectrum of states and studying their relationships,
both between states with the same quark content but different quantum
numbers, and vice versa.  The hadrons containing strange quarks are
particularly interesting to study because they lie in a middle ground
between the nearly massless up \& down quarks and the heavy charm \&
bottom quarks, and while a rich spectrum of strange quark hadrons is
predicted comparatively very little is known about them. Over the past
two decades, meson photo- and electroproduction data of unprecedented
quality and quantity have been measured at facilities worldwide,
leading to a revolution in our understanding of baryons consisting of
the lightest quarks, while the corresponding meson beam data are mostly
outdated or non-existent~\cite{meson_beams}.  For the study of strange quark hadrons, a kaon beam~\cite{KL2016_proc} has the advantage over photon or pions beams of having strange quarks in the initial state, which leads to enhanced production of
these states.  A secondary $K_L$ beam provides a unique probe for such
studies, and by using a primary high intensity photon beam, a
high-quality $K_L$ beam with low neutron background can be generated.
In conjunction with a large acceptance spectrometer, this enables the
measurement of cross sections and polarizations of a range of hyperon
production reactions, and allows for the identification of the quantum
numbers of identified states and provides the opportunity to make a
similar leap forward in our understanding of the strange hadron
spectrum.  The study of strange hadron spectroscopy using an intense
$K_L$ beam is the topic of Jefferson Lab Experiment C12-19-001~\cite{Klong}.

\section{Science Method}
\label{sec:science-method}

\begin{figure}
\includegraphics[width=3.0in]{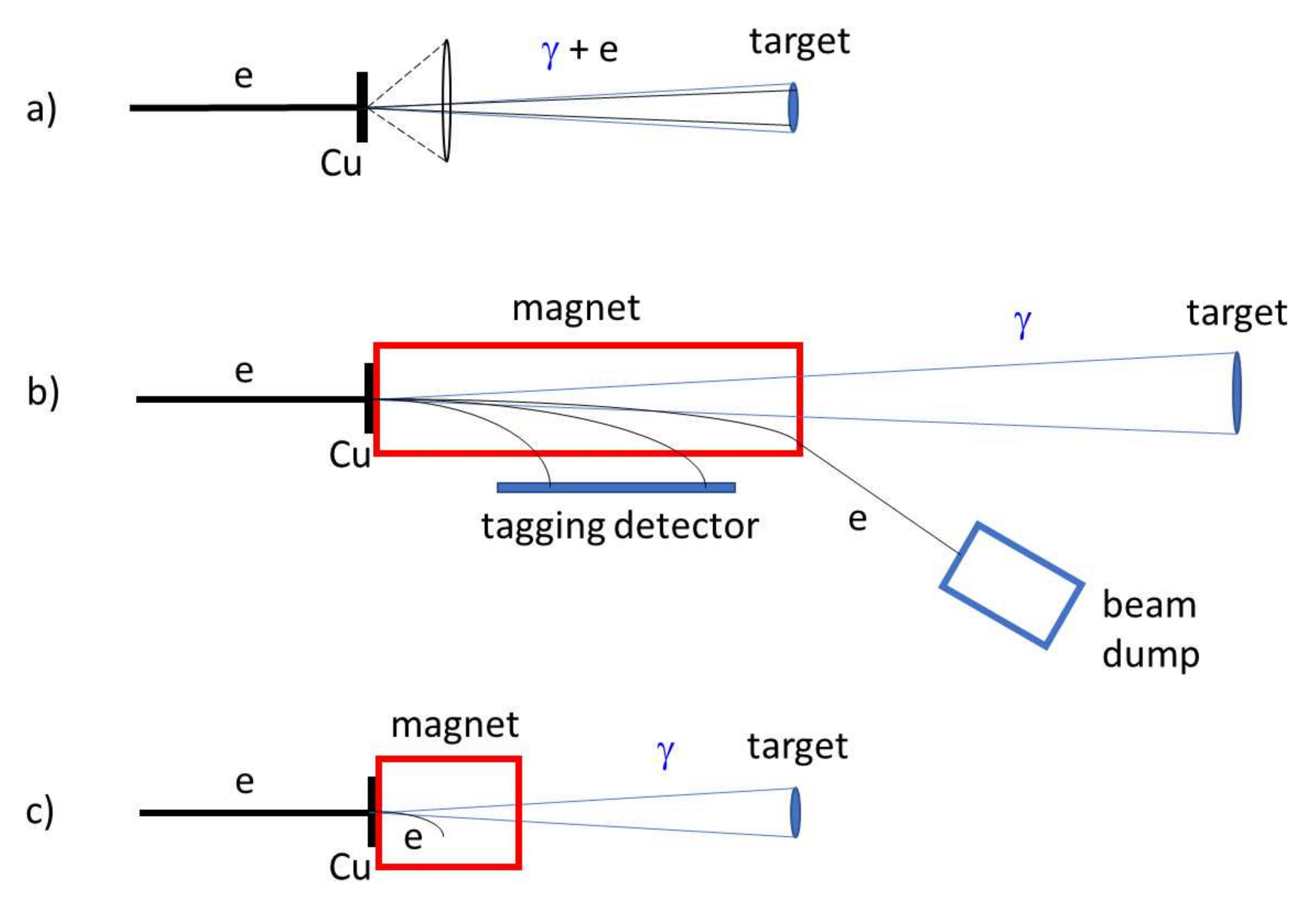}
\caption{\label{fig-photon-sources} \it Different schemes to produce
  high-energy photon beams. Scheme a) is the traditional
  bremsstrahlung technique where a copper radiator is placed in an
  electron beam resulting in a mixed photon and electron beam. In
  scheme b) a deflection magnet and beam dump are used to peel off the
  electrons and produce a photon-only beam. Scheme c) is the new CPS
  technique, with a compact hermetic magnet-electron dump and a narrow
  pure photon beam.}
\end{figure}

One of the traditional experimental techniques for producing a beam of
high-energy photons is to allow an electron beam to strike a radiator,
most commonly copper, producing a cone of bremsstrahlung photons which
are consequently mixed with the electron beam (see
Fig.~\ref{fig-photon-sources}a). The spread in the photon and outgoing
electron beams is dominated by electron multiple scattering, and for
electron beam energies of a few GeV is typically less than 1~mrad.
Accompanying this mixed photon and electron beam are secondary
particles produced in the electron-nuclei shower and characterized by
a much larger angular distribution (the extent of these secondary
cones are highlighted in the figure). For example, the cone of
secondary particles that survive filtering through a heavy absorber
material of one nuclear interaction length ($\approx$140-190~g/cm$^2$
or $\approx$15~cm) has an angular spread of 100-1000~mrad.  Although
this is the preferred technique for producing the largest flux of
photons, drawbacks include the fact that the beam is a mix of both
photons and electrons, that the photon beam energy is not a priori
known, and that the method is accompanied by the potential for large
radiation background dose due to the large spread of secondary
particles produced.

An alternative technique for producing a photon beam involves the use of
a radiator, a deflection magnet and a beam dump for the undeflected
electrons, augmented for energy-tagged photon beams with a set of
focal plane detectors covering a modest to large momentum acceptance
(see Fig.~\ref{fig-photon-sources}b). A configuration like this
requires significant space along the beam direction and heavy
shielding around the magnet and the beam dump, which have large
openings due to the large angular and energy spread of the electrons
after interactions in the radiator. In addition, without tight
collimation the traditional scheme leads to a large transverse size of
the photon beam at the target due to divergence of the photon beam and
the long path from the radiator to the target. This can be an issue as
the beam spot size contributes to the angular and momentum
reconstruction resolution of the resultant reaction products due to
uncertainty in the transverse vertex position. The advantage of this
method is that one has a pure photon beam, and if augmented with a set
of focal-plane tagging detectors the exact photon energies can be
determined. A significant drawback is that in order to keep
focal-plane detector singles rates at a manageable level (typically
less than a few MHz) the flux of incident electrons must be modest
($\approx$ 100~nA) and, correspondingly, the photon flux is less than
might otherwise be possible.

The proposed CPS concept (see Fig.~\ref{fig-photon-sources}c)
addresses the shortcomings of these two traditional widely-used
experimental techniques. The concept takes advantage of the modest
spread of the photon beam relative to the angular distribution of the
secondary particles produced in the electron-nuclei shower. It does so
by combining in a single shielded assembly all elements necessary for
the production of the intense photon beam and ensures that the
operational radiation dose rates around it are acceptable (see
Ref.~\cite{CPS-Concept}). Much of this is achieved by keeping the
overall dimensions of the apparatus limited, and by careful choice and
placement of materials.

The CPS conceptual design features a magnet, a central copper absorber
to handle the power deposition, and tungsten powder and borated
plastic to hermetically shield the induced radiation dose as close to
the source as possible. The magnet acts as dump for the electrons with
a cone of photons escaping through a small collimator.  The size of
the collimator can be chosen to be as narrow as the photon beam size,
taking into account natural divergence plus the size of the electron
beam raster. The concept of a combined magnet-dump allows us to reduce
dramatically the magnet aperture and length, as well as the weight of
the radiation shield, due to the compactness and hermeticity (with
minimized openings) of the system, thus significantly reducing the
radiation leakage. This conceptual approach opens a practical way
forward for a CPS, providing one can manage both the radiation
environment in the magnet and the power deposition density in the
copper absorber.

\begin{figure}
\includegraphics[width=3.0in]{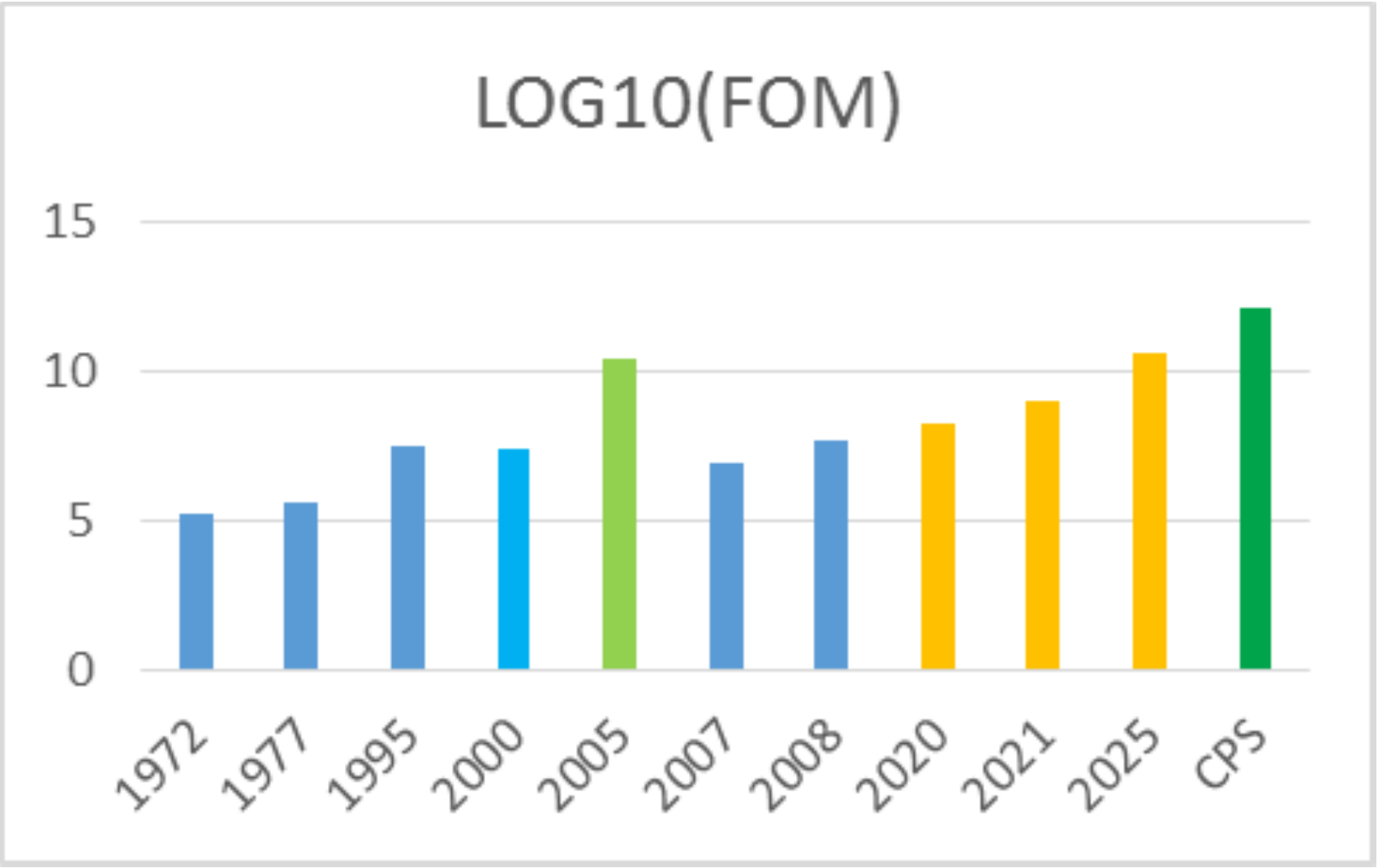}
\caption{\label{fig-cps-fom} \it The figure-of-merit (FOM) of photon
  beam experiments with dynamically nuclear polarized targets, defined
  as the logarithm of the effective photon beam intensity multiplied
  by the averaged target polarization squared, as a function of time.
  Note the large gain enabled by the CPS.  The indicated FOM in 1972,
  1977, 1995, 2007 and 2008 are based on actual experiments at
  Daresbury, Bonn, Jefferson Lab and
  Mainz~\cite{Court1980,Crabb2997,Others}. The FOM noted in 2000 and
  2005 are based upon proposed setups at SLAC and Jefferson Lab, with
  the latter closest in concept to the CPS. We also add the projected
  FOM of approved future experiments at HiGS/Duke and Jefferson Lab.}
\end{figure}

Compared to the more traditional bremsstrahlung photon sources
(Figs.~\ref{fig-photon-sources}a and \ref{fig-photon-sources}b and
{\sl e.g.}~Refs.~\cite{Anderson1970,Tait1969}), the proposed solution
offers several advantages, including an intense and narrow pure photon
beam and much lower radiation levels, both prompt and post-operational
from radio-activation of the beam line elements.  The drawbacks are a
somewhat reduced photon flux as compared to the scheme of
Fig.~\ref{fig-photon-sources}(a), and not having the ability to
directly measure the photon energy as in the scheme of
Fig.~\ref{fig-photon-sources}(b).

The primary gain of the CPS, and the reason for much of the initial
motivation, is for experiments using dynamically nuclear polarized
(DNP) targets, with an estimated gain in figure-of-merit of a factor
of 30 (see Fig.~\ref{fig-cps-fom}).  Dynamic nuclear polarization is
an effective technique to produce polarized protons, whereby a
material containing a large fraction of protons is cooled to low
temperatures, $<$1~K, and placed in a strong magnetic field, typically
about 5~Tesla~\cite{Averett99,Pierce14}.  The material is first doped,
either chemically or through irradiation, to introduce free radicals
(electrons).  The low-temperature and high-field conditions cause the
electrons to self-polarize, and their polarization is then transferred
to the proton using microwave techniques. These conditions however
impose a serious limitation: beams traversing the polarized target
material will produce ionization energy losses that simultaneously
heat and depolarize the target. They also produce other harmful free
radicals which allow further pathways for proton polarization to
decay.  This limits the local beam intensities the polarized target
material can handle.

Conventional target cells have diameters much larger than the
desirable beam spot size, and one is forced to minimize rapid
degradation of the target polarization by the beam at one location at
the target.  The traditional solution of minimizing such localized
polarization degradation is fast movement of the beam spot, which
allows avoiding overheating of the material and ensuring that the
depolarizing effects of the beam are uniformly spread over the target
volume.

A beam raster magnet, which moves the beam with a frequency of several
Hz, was used in past experiments in Jefferson
Lab~\cite{Averett99,Zhu01,Pierce14}.  However, this does not work for
very small collimation apertures, e.g.~a few mm by a few mm
collimation cone, limiting possible beam motion. The CPS solution for
the beam-target raster thus includes a combination of the target
rotation around the horizontal axis and $\pm$10~mm vertical motion of
the target ladder. Such a raster method effectively moves the motion
complexity out of the high radiation area of the absorber.  The same
effect can be achieved by vertical displacement of the beam spot,
i.e.~by a small variation of the vertical incident angle of the
electron beam at the radiator. With a $\pm$5~mrad vertical angle
variation and 200~cm distance between the radiator and the target, the
displacement of the beam spot is equal to $\pm$1~cm, about the size of
the conventional target cells.

Traditionally, such photon beam experiments have been performed using
the scheme indicated in Fig.~\ref{fig-photon-sources}a. This limits
the electron beam current to less than 100~nA to prevent rapid target
polarization damage. With the CPS scheme, we anticipate use of an
electron beam current of up to 2.7~$\mu$A to provide the photon flux
for an equivalent heat load in the DNP target. Hence, we gain a factor
of about 30. The history of the figure-of-merit of bremsstrahlung
photon beam experiments with DNP targets is further illustrated in
Fig.~\ref{fig-cps-fom}.

\section{The Compact Photon Source - Description of Instrumentation}
\label{sec:cps}

 The physics program described above requires a high-intensity and
 narrow polarized photon beam and a polarized target to access the
 exclusive photoproduction reactions in order to extract the relevant
 experimental observables.  The CPS provides a compact solution with a
 photon flux of $1.5 \times 10^{12}$ equivalent photons/s.

\subsection{Conceptual Design}
\label{sec:concept}

\begin{figure}
\centering
\includegraphics[width=3.0in]{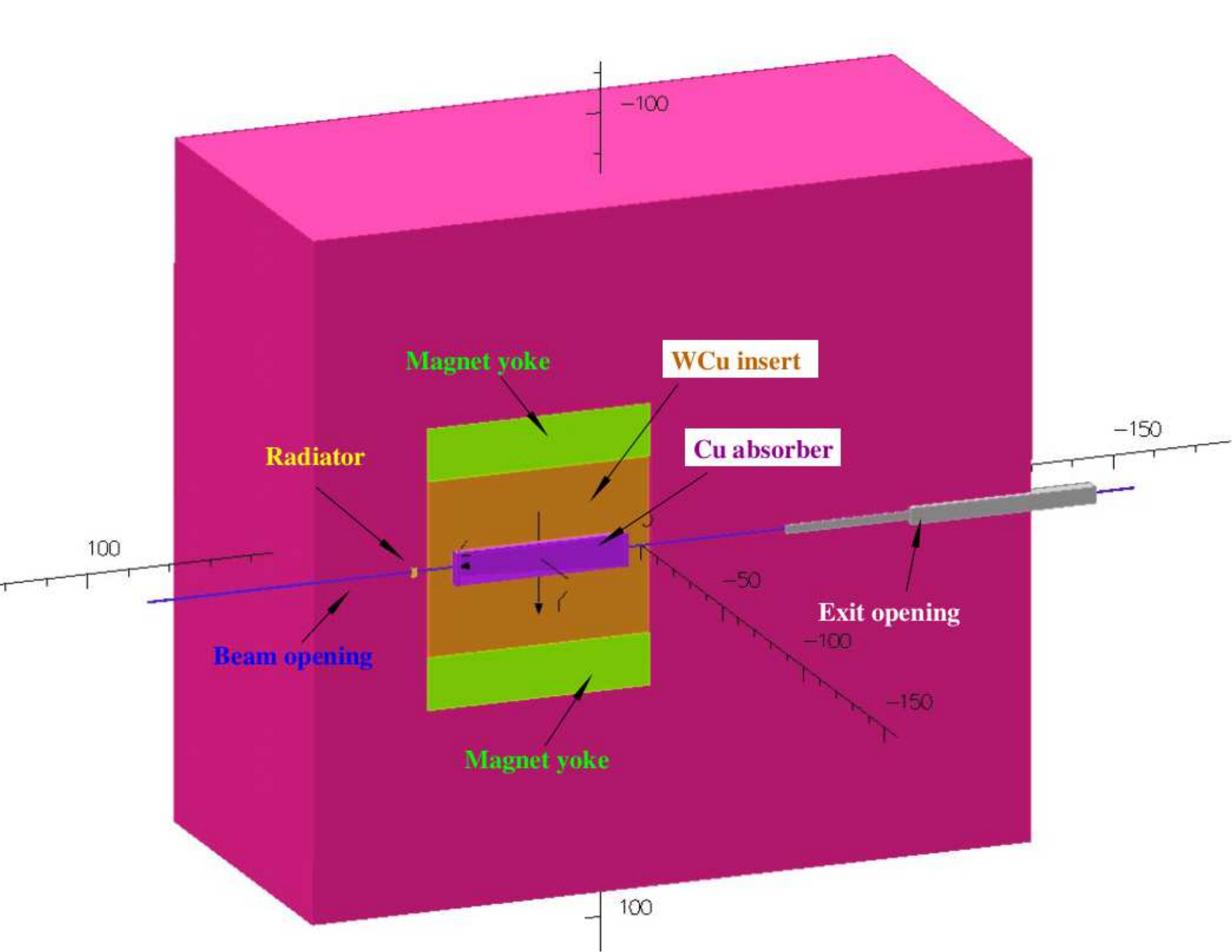}
\caption{\label{fig:CPS} \it The CPS cut-out side view. Deflected electrons strike
          a copper absorber, surrounded by a W-Cu insert inside the
          magnet yoke. The outer rectangular region in this view is
          the tungsten-powder shield.}
\end{figure}

        The main elements of the CPS are shown in Fig.~\ref{fig:CPS}.
        Without loss of photon intensity, a channel (a collimator for
        the secondary radiation) around the photon beam can be as
        narrow as the photon beam size. After passing through the
        radiator, the electron beam should be separated from the
        photon beam by means of deflection in a magnetic field.  The
        length, aperture and field strength of the magnet are very
        different in the proposed source compared to in the
        traditional tagging technique.  In the traditional source the
        magnet is needed to direct the electrons to the dump.  Because
        of the large momentum spread of electrons which have
        interacted in the radiator, the magnet aperture needs to be
        large and the dump entrance even larger: 13\% of the beam
        power is therefore lost before the beam dump, even with a 10\%
        momentum acceptance of the beam line.  In contrast, in the
        proposed source the magnet acts as dump for the electrons with
        a cone of photons escaping through a small collimator.

\begin{figure}
\centering
\includegraphics[width=3.0in]{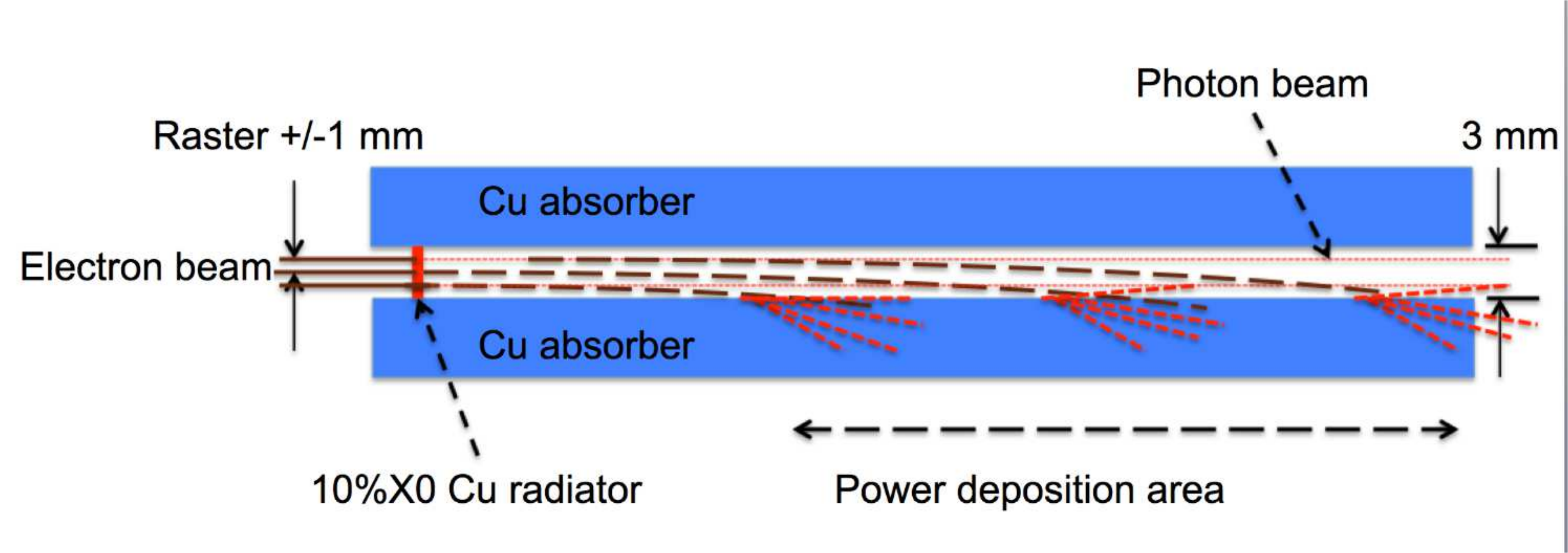}
\caption{\label{fig:beam} \it The scheme of beam deflection in the magnetic field
          to the absorber/dump.}
\end{figure}

        The dumping of the electron beam starts in the photon beam
        channel, so even a small deflection of the electron trajectory
        by just 1-3~mm due to the presence of the magnetic field is
        already sufficient to induce a shower.  At the same time, such
        a deflection needs to be accomplished at a relatively short
        distance (much shorter than the size of the radiation
        shielding) after the beam passes through the radiator to keep
        the source compact.  Indeed, 
        in the proposed CPS magnet design the trajectory radius is
        about 10~m for 11~GeV electrons, the channel size is 0.3~cm,
        and the raster size is 0.2~cm, so the mean distance travelled
        by an electron in the magnetic field is around 17~cm, with a
        spread of around 12~cm (see the scheme in
        Fig.~\ref{fig:beam}).  Therefore, a total field integral of
        1000~kG-cm is adequate for our case, which requires a 50~cm
        long iron-dominated magnet.

\subsection{Magnet}
\label{sec:magnet}

\begin{figure}
\centering
\includegraphics[width=3.0in]{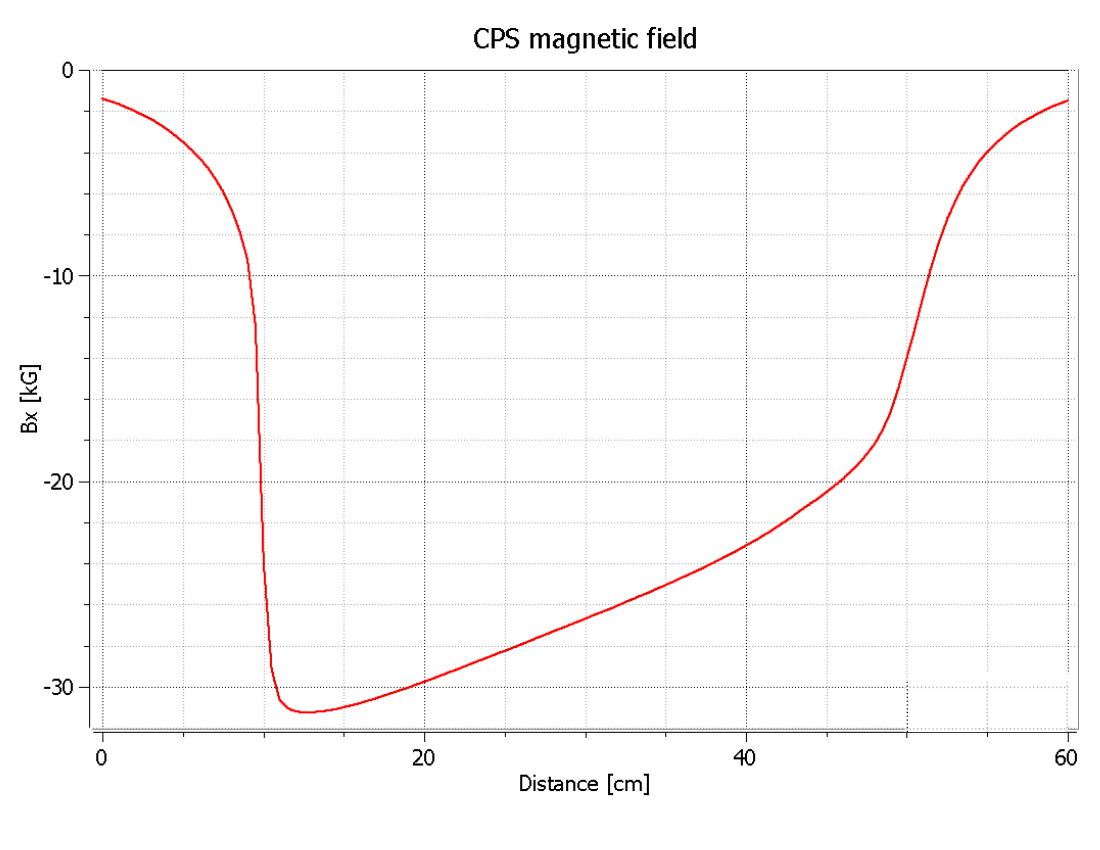}
\caption{\label{fig:field} \it Magnetic field ($B_x$) profile along the beam
          direction, as a function of distance from the radiator
          position.}
\end{figure}

Normal conducting magnets for operation in high levels of radiation
have been constructed at several hadron facilities, including the
neutron spallation source at ORNL and the proton complex
JPARC~\cite{Tanaka11,Petrov16}.  The magnet designed for the CPS has
permendur poles tapered in two dimensions, which allows for a strong
magnetic field at the upstream end of the magnet (3.2~T), with the
coils located 20~cm from the source of radiation.  The resulting
radiation level at the coil location was calculated to be sufficiently
low (below 1~Mrem/hr) to allow the use of relatively inexpensive
kapton tape based insulation of the coils~\cite{Kapton}. As discussed
above, the length of the magnet was selected to be 50~cm and the field
integral 1000~kG-cm.  Fig.~\ref{fig:field} shows the longitudinal
profile of the magnetic field obtained from OPERA calculations.

\subsection{Central Absorber}
\label{sec:absorber}

\begin{figure}
\centering
\includegraphics[width=3.0in]{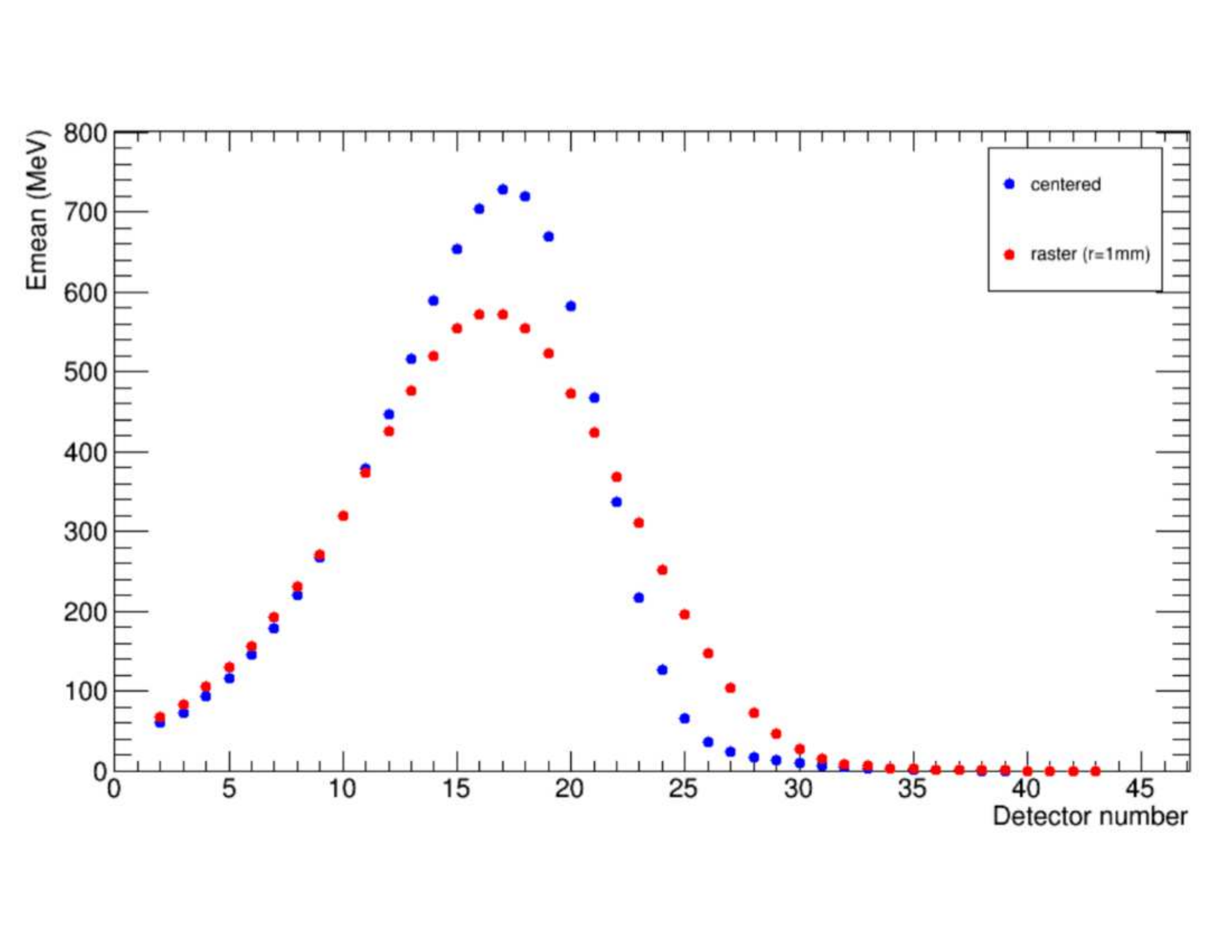}
\caption{\label{fig:power} \it Longitudinal profile of the energy distribution
          (integrated for one cm copper slab) for a 11 GeV incident
          electron beam.  The maximum power density occurs at a
          distance of 18~cm from the radiator. The blue dots show the
          energy deposition for the electron beam centered in a 3~mm
          by 3~mm channel, while the red dots show the same for the
          beam rastered with a radius of 1~mm.}
\end{figure}

The beam power from the deflected electron beam and subsequent shower
is deposited in an absorber made of copper, whose high heat
conductivity helps to manage the power density.  An absorber made of
aluminum would help to reduce power density by a factor of 2-3
compared with copper due to its smaller radiation length, but it would
also increase the length of the CPS by about 50~cm so is not
preferred.  The heat removal from the copper absorber is arranged via
heat conduction to the wider area where water cooling tubes are
located.  Fig.~\ref{fig:power} shows the simulated longitudinal
profile of the power density.
%\vskip .3 in

\begin{figure}
\centering
\includegraphics[width=3.5in]{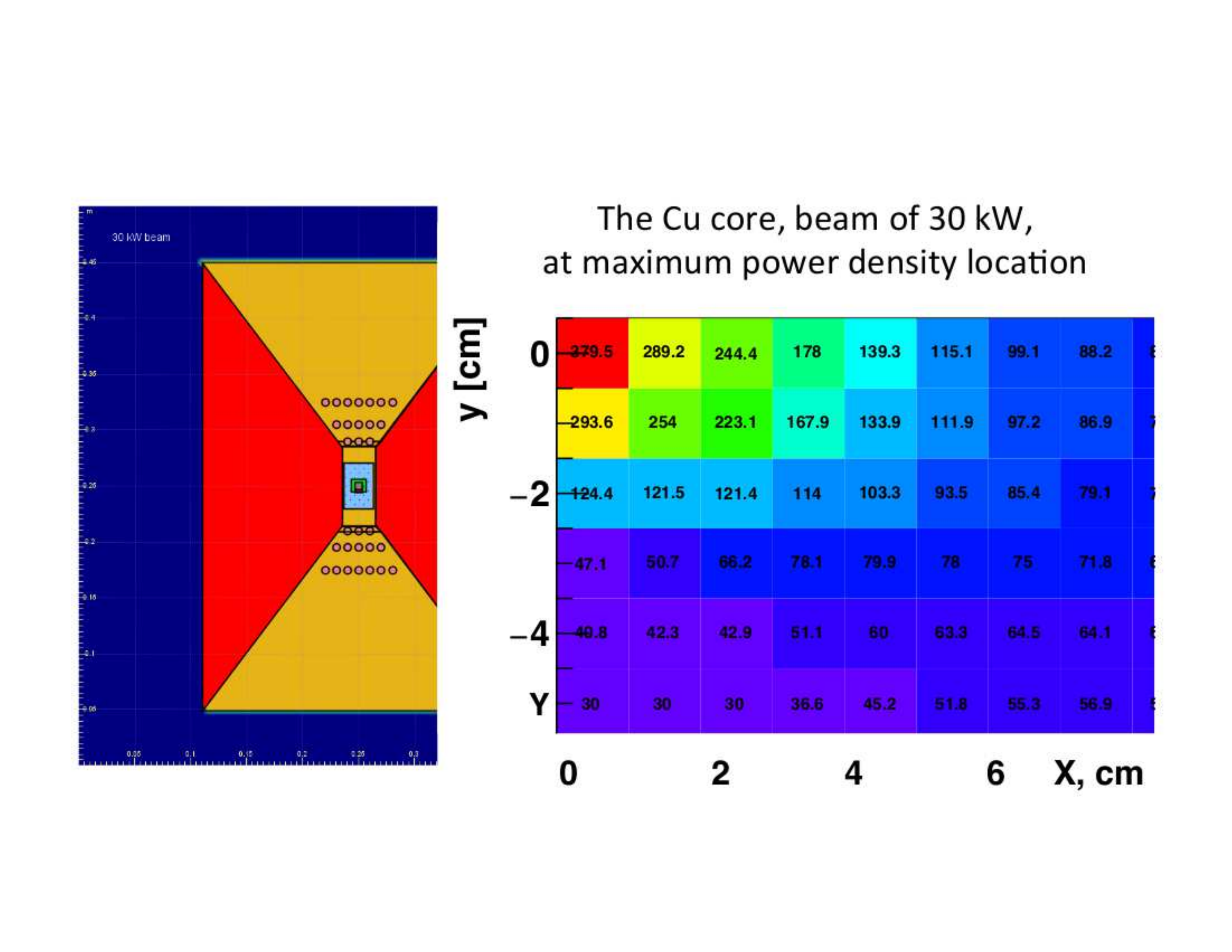}
\caption{\label{fig:temperature} \it Left panel: the cross section of the absorber with
          the water cooling channels (the copper is shown in light
          blue and the W-Cu(20\%) is shown in gold).  Right panel: the
          temperature map for 1 cm by 1 cm elements at the
          longitudinal coordinate of the power deposition maximum.}
\end{figure}

        The transverse distribution of power is also very important to
        take into account because, for a high energy incident beam, it
        has a narrow peak.  Simulation of the deposited power density
        and 2-dimensional heat flow analysis were performed to
        evaluate the maximum temperature in the absorber.
        Fig.~\ref{fig:temperature} (left panel) shows the layout of
        materials in the model used for the temperature analysis.  The
        calculation was performed for an 11~GeV, 30~kW beam and a
        radiator with 10\% radiation length thickness. The resultant
        temperature was found to be below 400$^\circ$C, which is well
        in the acceptable range for copper. Fig.~\ref{fig:temperature}
        (right panel) shows the temperature profile in the transverse
        plane at the longitudinal location of maximum power
        deposition. Cooling of the core will require about four
        gallons of water per minute at 110~psi pressure (at
        30$^\circ$C temperature rise), which is easy to provide.

\subsection{Tungsten-powder Shield}
\label{cps-w-powder-shield}

\begin{figure}
\centering
\includegraphics[width=3.0in]{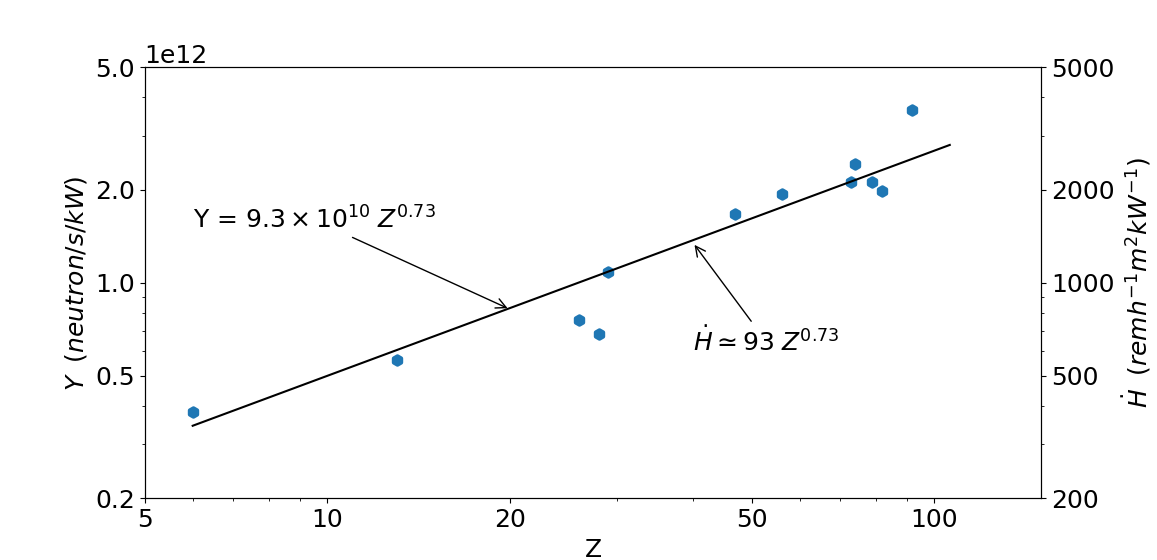}
\caption{\label{fig:n-yield} \it The neutron yield and dose rate for a 500~MeV
          incident electron beam as a function of atomic number (based
          on an original figure from SLAC~\cite{Swanson77}).}
\end{figure}

The amount of material needed for radiation shielding is primarily
defined by the neutron attenuation length, which is 30~g/cm$^2$ for
neutrons with energy below 20~MeV and 125~g/cm$^2$ for high energy
neutrons.  The neutron production rate by an electron beam in copper
is $1\times 10^{12}$ per kW of beam power according to
Ref.~\cite{Swanson77} (see Fig.~\ref{fig:n-yield}).  At a distance of
16~meters from the unshielded source for a 30~kW beam, the neutron
flux would be $1\times 10^7$~n/cm$^2$/s, which would produce a
radiation level of 110~rem/hr.  The proposed conceptual design has a
total shield mass of 850~g/cm$^2$ and will result in a reduction in
these radiation levels by a factor of around 1000.

        The space inside the magnet between the poles and coils is
        filled by an inner copper absorber and an outer W-Cu(20\%)
        insert, which provides a good balance between effective beam
        power absorption and radiation shielding.  For the shield
        outside the magnet, the current design employs tungsten
        powder, whose high density (16.3~g/cm$^3$) \footnote{The
          density of tungsten is 19.25~g/cm$^3$, but more commonly
          admixtures of tungsten and Cu/Ni, or in this case tungsten
          powder, are used with somewhat lower densities} helps to
        reduce the total weight of the device.  A thickness of 50~cm
        was used as a first iteration for the thickness of the outer
        shield of the CPS, but we have investigated the impact of
        varying this amount of outer shielding and adding borated
        plastic (as discussed later).

\subsection{Impact on Polarized Target}
\label{cps-target-field}

The most significant gain associated with deployment of the CPS is for
experiments using dynamically polarized targets. However, such
polarized targets operate with strong polarizing fields themselves.
In addition, dynamically polarized target operation imposes strict
requirements on the field quality at the target location, where fields
and gradients need to be compensated at the 10$^{-4}$ level. This
necessitates studies of the mutual forces associated with the
2-3~Tesla CPS dipole magnet and the 5~Tesla polarized target solenoid,
in terms of both the design of the support structures and the
experimental operation.

\begin{figure}
\centering
\includegraphics[width=3.0in]{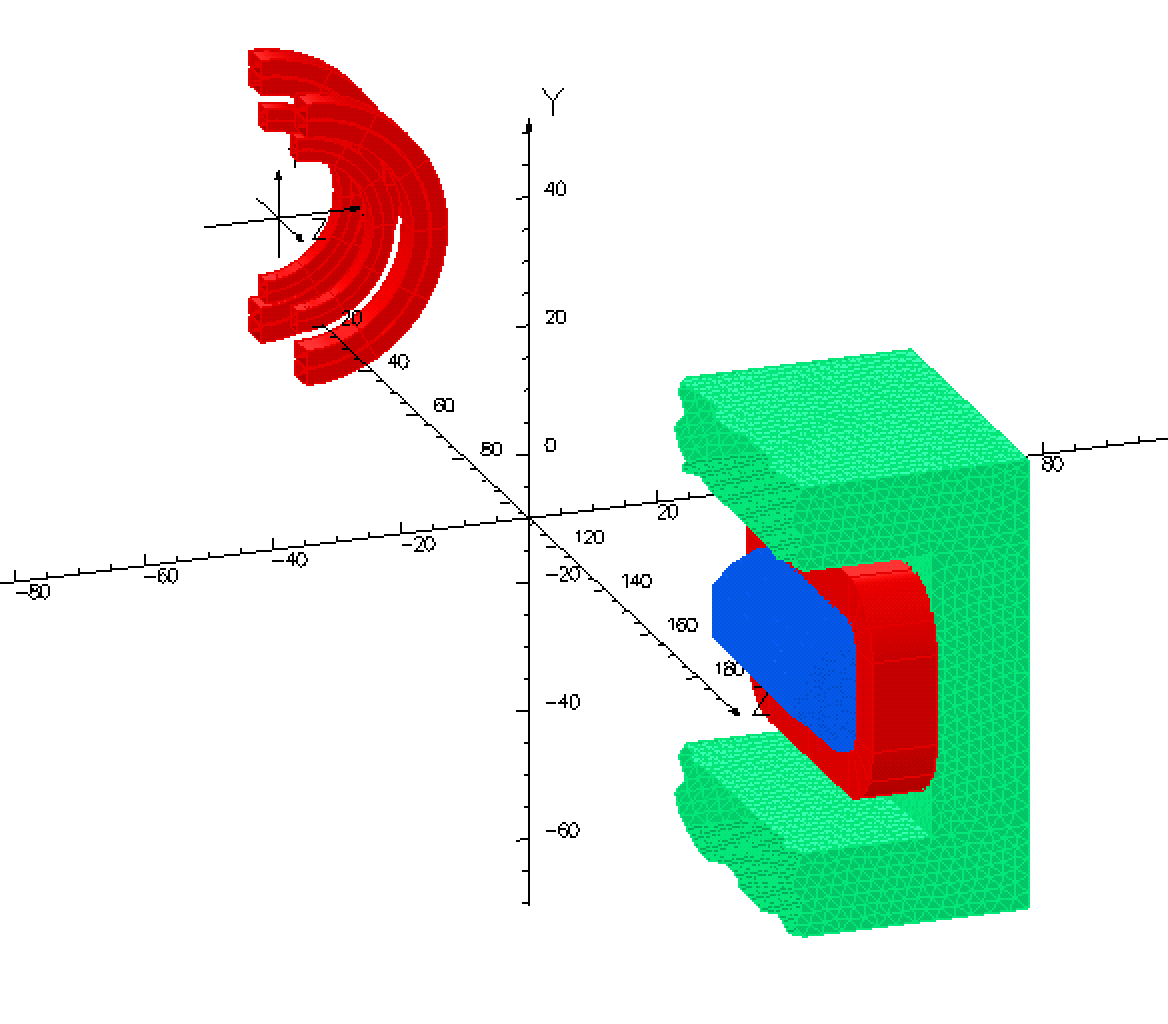}
\caption{\label{fig:modelT} \it The TOSCA model used in the field and force calculations for longitudinal orientation of the coils/target polarization.}
\end{figure}

\begin{figure}
\centering
\includegraphics[width=3.0in]{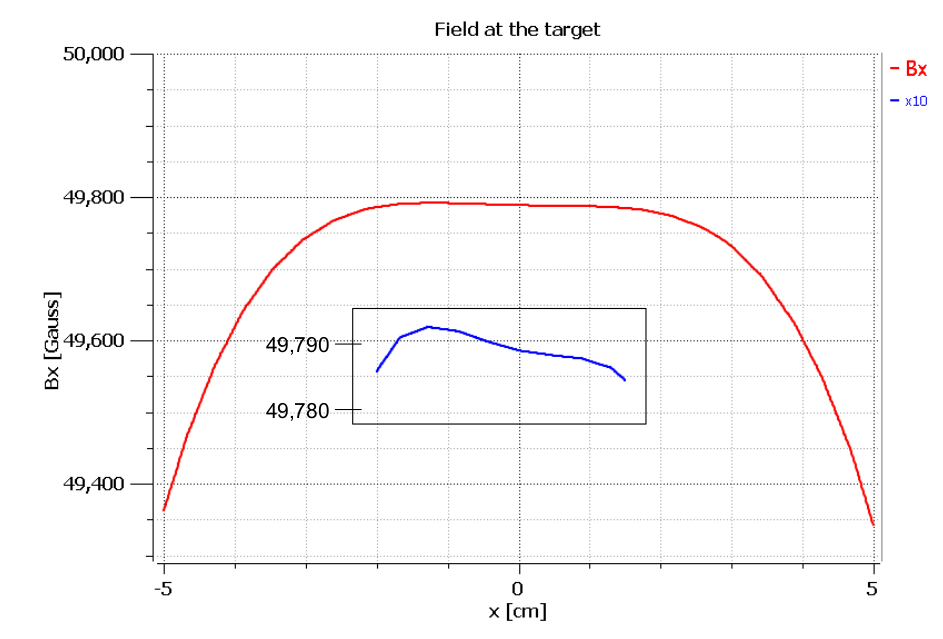}
\caption{\label{fig:gradient} \it The field at the target center. The insert shows the field zoomed by a factor of 10.}
\end{figure}

        The fields associated with the combination of these two
        magnetic systems were calculated using the model shown in
        Fig.~\ref{fig:modelT}, with the following
        results obtained:
        \begin{itemize}
        \item When the CPS is on but the polarized target magnet is
          off, the (total) field at the target location is 0.1~Gauss.
        \item When the polarized target magnet is on and the CPS is
          off or removed, the field at the CPS location is about
          130~Gauss.
        \item When both the CPS and the polarized target magnet
          are ON, the field gradient at the polarized target center is
          about 2 Gauss/cm (Fig.~\ref{fig:gradient}).
        \end{itemize}

        These results show that, for the CPS the induced field is
        mainly due to the CPS magnet yoke becoming polarized by the
        target field.  Whereas for the target, the field gradient at
        the target location is sufficiently low for routine
        dynamically polarized NH$_3$ or ND$_3$ operation, with a
        relative values of around $0.4 \times 10^{-4}$.

\section{Radiation Requirements}
\label{sec:radiation-requirements}

As discussed previously, most of the proposed Jefferson Lab
experiments with the CPS will utilize a dynamically nuclear polarized
target.  Electron beam currents for use with such targets are
typically limited to 100~nA or less, to reduce both heat load and
radiation damage effects.  The equivalent heat load for a pure photon
beam impinging on such a target corresponds to a photon flux
originating from a 2.7~ $\mu$A electron current striking a 10$\%$
copper radiator. The radiation calculations presented in this section
therefore assume a CPS able to absorb 30~kW of beam power
(corresponding to a beam of 11 GeV electrons with a current of 2.7
$\mu$A). In addition, the beam time assumed for a typical experiment
is 1000 hours.

For such an experiment at Jefferson Lab, the following radiation
requirements must be fulfilled:
\begin{itemize}
\item{The prompt dose rate in the experimental hall must be $\le$ several
    rem/hr at a distance of 30 feet from the CPS.}
\item{The activation dose outside the CPS envelope at a distance of
    one foot must be $\le$ several mrem/hr one hour after the end of a
    1000 hour run.}
\item{The activation dose at the centre of the experimental target
    area, where operational maintenance tasks may be required at a
    distance of one foot from the scattering chamber must be $\le$
    several mrem/hr one hour after the end of a 1000 hour run.}
\end{itemize}

The CPS conceptual design has been established with the aid of several
extensive simulations. As validation of the simulation tools used,
benchmark comparisons were made with GEANT3, GEANT4, FLUKA and DINREG.
~\cite{Geant4,Fluka}\footnote{Note that these codes calculate
  particle yields/s/cm$^2$, which have to be converted into the
  effective dose rate (in rem/hr) using Fluence-to-Effective Dose
  conversion factors~\cite{ICRP116} taking into account an
  energy-dependence factor.}. After benchmark validation, a series of
radiation calculations were performed in order to:
\begin{itemize}
\item{Determine the size and layout of the shielding around the
    magnet, and the choice of materials (copper, Cu-W alloy, concrete,
    borated plastic, etc.).}
\item{Determine the magnet field requirements in terms of peak field,
    gap size, and field length.}
\item{Determine the radiation levels on the magnet coils, and based on
    these results to identify radiation hardened materials that might
    be used in building the coils.}
\item{Determine the radiation levels on the polarized target
    electronics.}
\item{Determine the radiation levels directly adjacent to the CPS as
    well as at the experimental hall boundary.}
\end{itemize}

\section{Radiation Studies and Shielding Design}
\label{sec:radstudies-shieldingdesign}

In this section we will describe studies performed for several
different experimental configurations in order to identify the various
sources of radiation and make direct comparisons of the calculated
dose rates. 

\subsection{Prompt Radiation Dose Rates}

In order to provide a baseline the prompt radiation dose originating
from a 2.7~$\mu$A electron beam hitting a 10\% copper radiator located
at a distance of 2.15~m upstream of the centre of the experimental
target was calculated.  As the geometry of the target system and CPS
are not included in this simulation, all prompt radiation originates
from the interaction between the primary electron beam and the
radiator.  The prompt radiation dose is calculated by summing over all
azimuthal angles in a radial range between 5 and 10~cm from the beam
line.

\begin{figure*}
\centering
\includegraphics[width=5.0in]{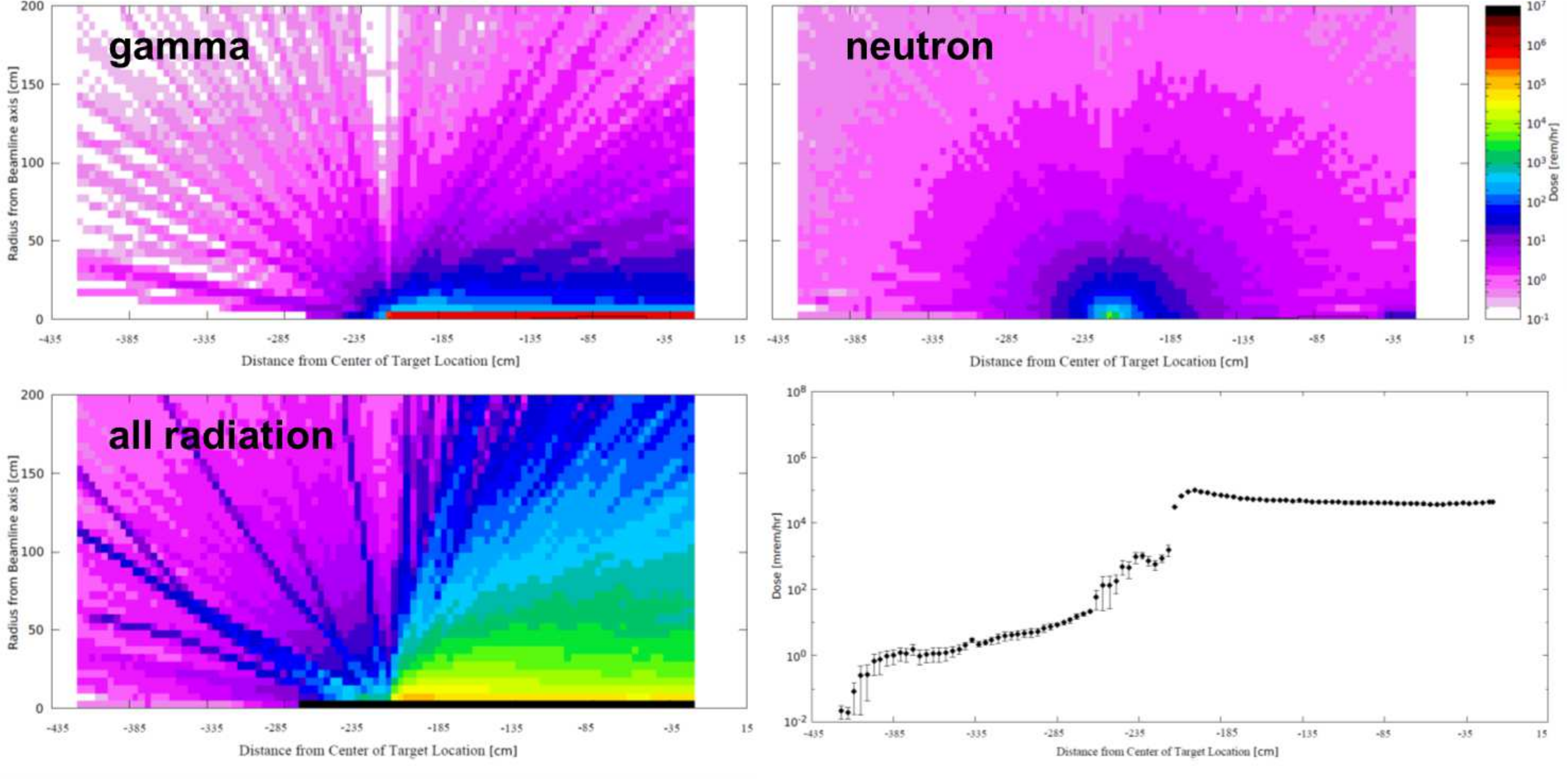}
\caption{\label{fig:prompt_no_cps} \it Prompt radiation dose rate
          as a function of position in the experimental hall for the
          case of a 2.7~$\mu$A electron beam hitting a 10\% copper
          radiator. Two-dimensional plots are shown for
          the dose from photons only (top left), from neutrons only
          (top right) and from all particle types (bottom left).  Also
          shown is a one-dimensional plot of prompt dose rate along
          the beam direction (bottom right).}
\end{figure*}

Fig.~\ref{fig:prompt_no_cps} shows two-dimensional dose rates
originating from photons only (top left), from neutrons only (top
right), from all particles (bottom left), and the one-dimensional
prompt radiation dose along the beam direction (bottom right).  With
the exception of the neutron contribution, most of the prompt
radiation is created along the beam direction, as expected. The prompt
radiation levels reach roughly 40~rem/hr, of which only around
200~mrem/hr is in the form of gamma radiation and 10~mrem/hr from
neutrons. The remaining and clearly dominant contribution is from
charged electron- and positron-induced showers.

\begin{figure*}
\centering
\includegraphics[width=5.0in]{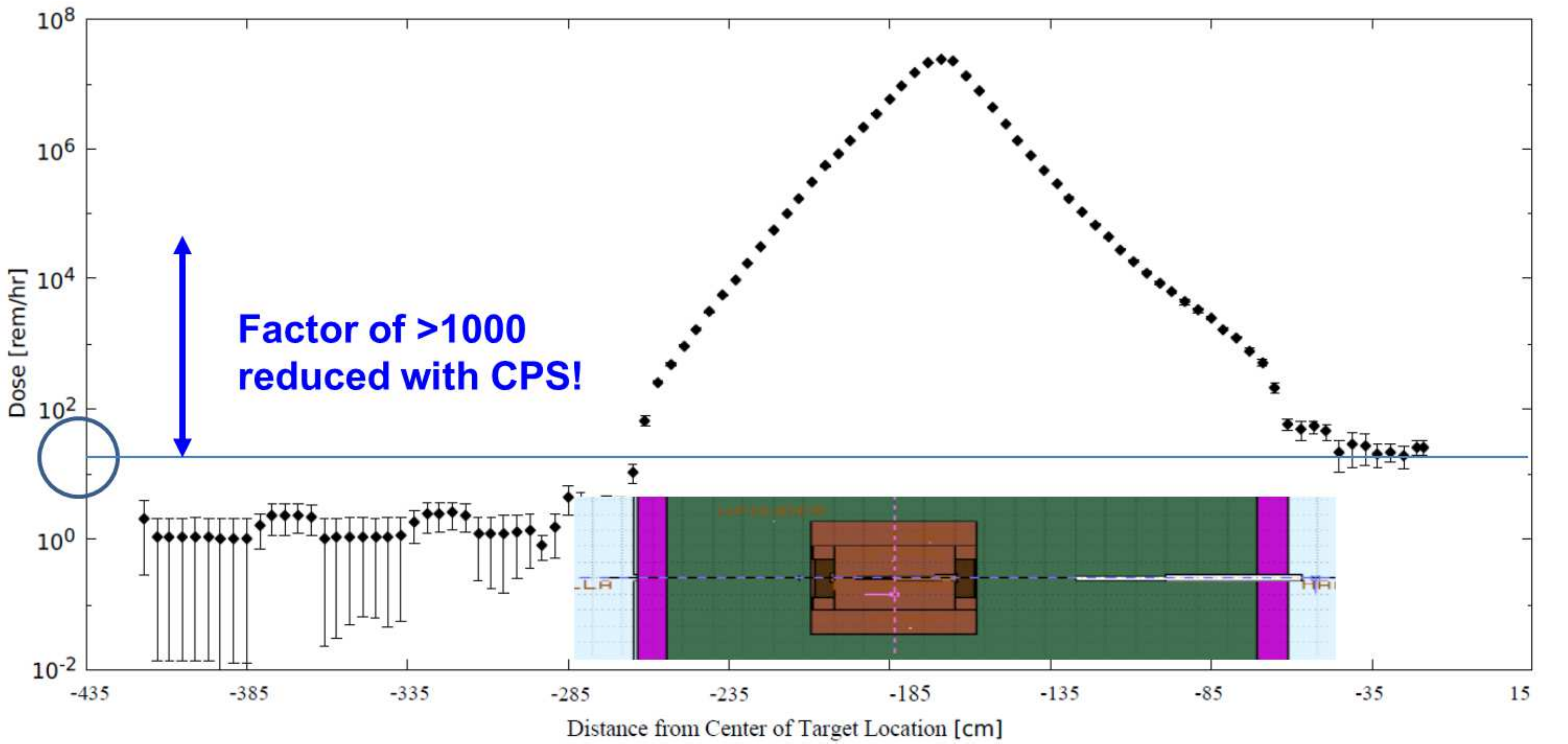}
\caption{\label{fig:prompt_cps} \it Prompt radiation dose rate as
          a function of upstream distance from the target for the case
          of a 2.7~$\mu$A electron beam hitting a 10\% copper radiator
          inside the CPS. The dose includes contirbtuions from all
          particles.  The large reduction factor of $>$1000 as a
          result of the CPS shielding is apparent.}
\end{figure*}

        The second scenario considered is that of a 2.7~$\mu$A
        electron beam incident on a 10\% copper radiator as before,
        but with the radiator located within the CPS geometry.
        Fig.~\ref{fig:prompt_cps} illustrates the prompt radiation
        dose along the beam direction for this case (note that the
        y-axis scale on this figure is the same as in
        Fig.~\ref{fig:prompt_no_cps}). One can clearly see that the
        prompt radiation levels within the CPS are much higher than
        before (around 300 times higher because the full power of the
        beam is now being deposited in the CPS).  Crucially, however,
        the prompt radiation dose rate outside the CPS is only around
        15~mrem/hr. Comparing this value for prompt dose rate to the
        one obtained above for the baseline scenario highlights the
        effect of the CPS shielding: there is a reduction by a factor
        of over 1000. This reduction is consistent with the factor
        estimated previously in section~\ref{cps-w-powder-shield}.

\begin{figure*}
\centering
\includegraphics[width=5.0in]{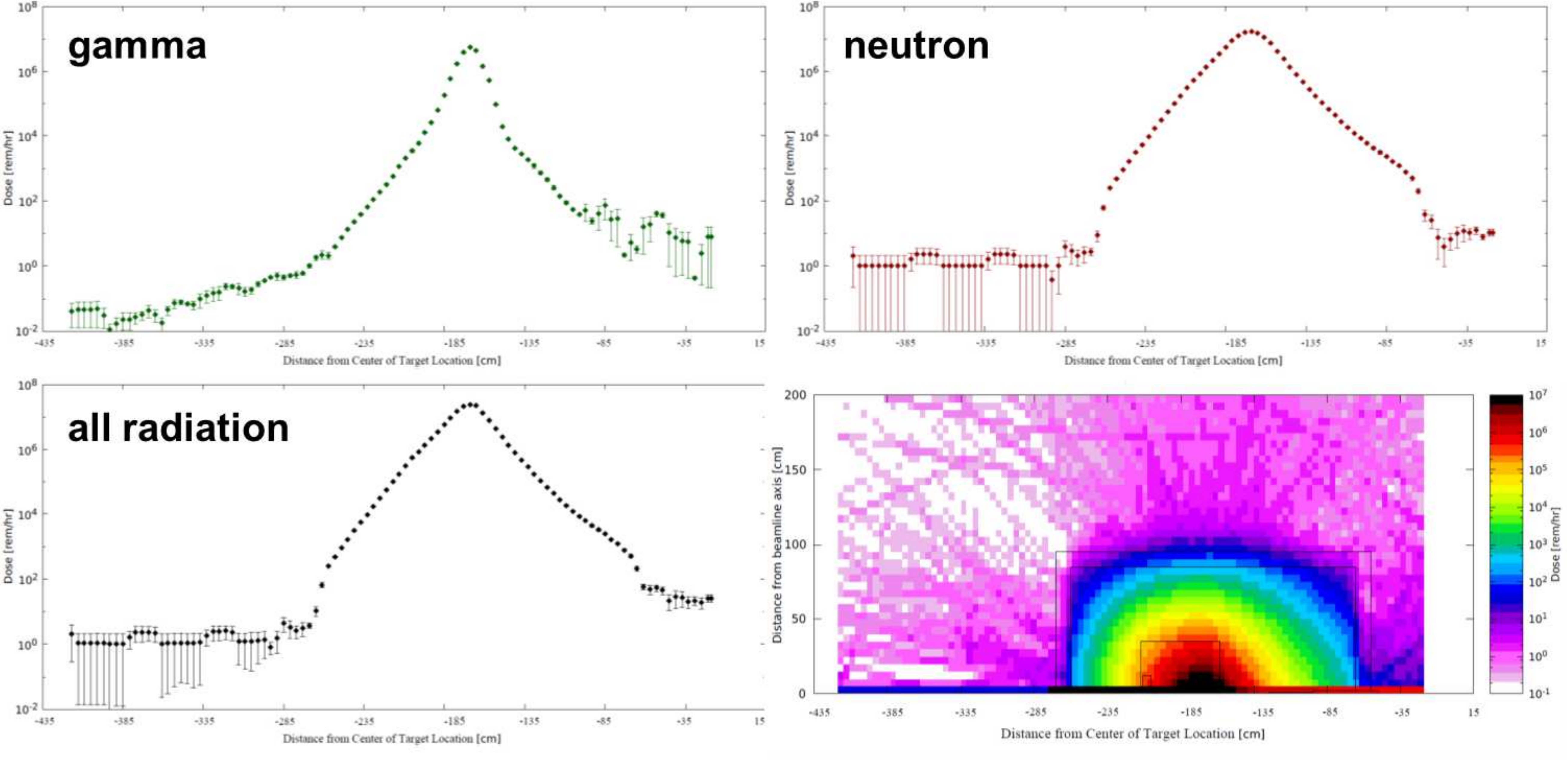}
\caption{\label{fig:prompt_cps-1D} \it Prompt radiation dose rate
          as a function of position in the experimental hall for the
          case of a 2.7~$\mu$A electron beam hitting a 10\% copper
          radiator inside the CPS. One-dimensional plots are shown for
          the dose from photons only (top left), from neutrons only
          (top right) and from all particle types (bottom left).  Also
          shown is a two-dimensional plot of prompt dose rate (bottom
          right), which shows the effectiveness of the CPS shielding
          concept.}
\end{figure*}

        This is a very important result, which is further illustrated
        in Fig.~\ref{fig:prompt_cps-1D}. In contrast with the baseline
        scenario, there are now no contributions to the overall prompt
        dose rate in the experimental hall from photons, electrons and
        positrons as these are all contained within the CPS shielding
        -- the neutron-only dose rate is nearly identical to the
        all-radiation rate. The bottom-right panel in
        Fig.~\ref{fig:prompt_cps-1D} illustrates how well optimized
        the CPS shielding concept is for absorbing prompt radiation.
        Outside the CPS the prompt radiation dose rate on the surface
        (indicated by the outer black rectangular lines on the figure)
        is reduced to a maximum level of roughly 10~rem/hr. This is
        due to the fact that the development of showers generated by
        interactions of the primary beam is highly suppressed and the
        resultant secondary charged particles and photons are fully
        contained.  This confirms that with a CPS the following
        requirement can be met: the prompt dose rate in the
        experimental hall $\le$ several~rem/hr at a distance of 30
        feet from the device.

\subsection{Impact of Boron and Shielding Optimization}

\begin{figure}
\centering
\includegraphics[width=3.0in]{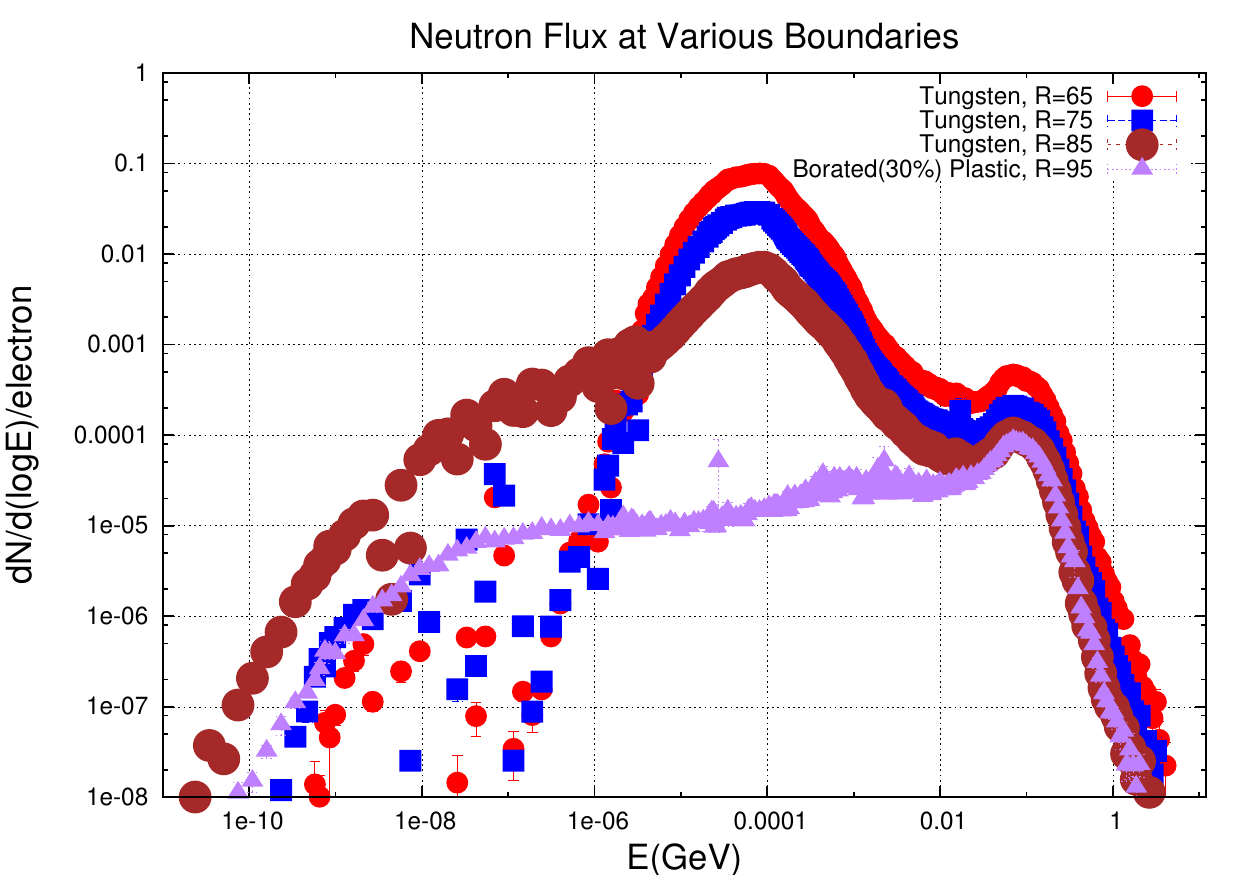}
\caption{\label{fig:boron} \it Neutron flux escaping the CPS for
          different shielding configurations, including the use of
          borated plastic.}
\end{figure}

It is well known that the neutron flux through a surface can be
drastically reduced by the addition of boron as a result of the very
high capture cross section of $^{10}$B. This effect was simulated by
calculating the neutron flux at the CPS boundary assuming various
thicknesses of tungsten shielding (65, 75 and 85~cm), and then adding
10~cm of borated (30\%) plastic. The result can be seen in
Fig.~\ref{fig:boron}, which shows the neutron flux as function of
neutron energy. Increasing the tungsten thickness clearly reduces the
neutron flux as expected, but a much more drastic reduction is seen
when the 10~cm of borated plastic is added. Thus, the baseline
conceptual shielding design of the CPS is assumed to be 85~cm thick
tungsten surrounded by 10~cm of borated plastic.

\begin{figure}
\centering
\includegraphics[width=3.0in]{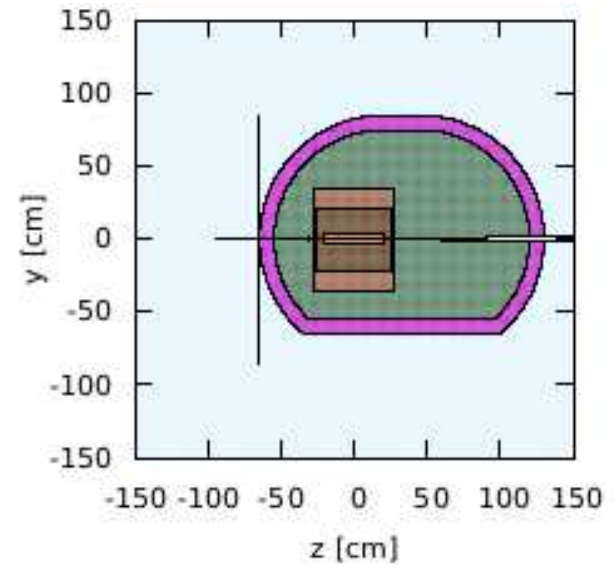}
\caption{\label{fig:FLUKA-oval} \it An alternative shielding
          design used in FLUKA radiation calculations with reduced
          W-powder overall, on the bottom-side and with rounded
          corners.}
\end{figure}

        The outer dimension of the tungsten-powder shielding as
        outlined for optimized shielding above is 1.7~m by 1.7~m by
        1.95~m, or a volume of 5.63~m$^3$.  One needs to subtract from
        this total volume the inner box including the magnet, which
        amounts to 0.26~m$^3$, leaving a net volume of 5.37~m$^3$, or
        88~tons of W-powder. There are various options to reduce the
        weight and therefore cost, if needed.  One could reduce the
        overall size of the W-powder shielding by 5~cm on each side.
        This would result in a reduction of the shield weight to 73
        tons, but would also lead to an increase of the radiation
        levels by about 50\%. If one would remove an additional 10~cm
        only on the bottom side, there would be a further increase of
        a factor of two in radiation level in the direction of the
        floor, but a further reduction in shielding weight to 68~tons.
        Alternatively, one could round the W-powder corners, as
        illustrated in Fig.~\ref{fig:FLUKA-oval}.  This would
        complicate modular construction, but would allow for similar
        radiation levels as with the optimized design, while reducing
        rhe shielding weight to $\approx$66 tons.

\subsection{Dose Rates due to Activation}

\begin{figure}[!htbp]
\centering
\subfigure[\label{fig:activation-cps} \it ]
% caption for subfigure a
{
\includegraphics[width=2.8in]{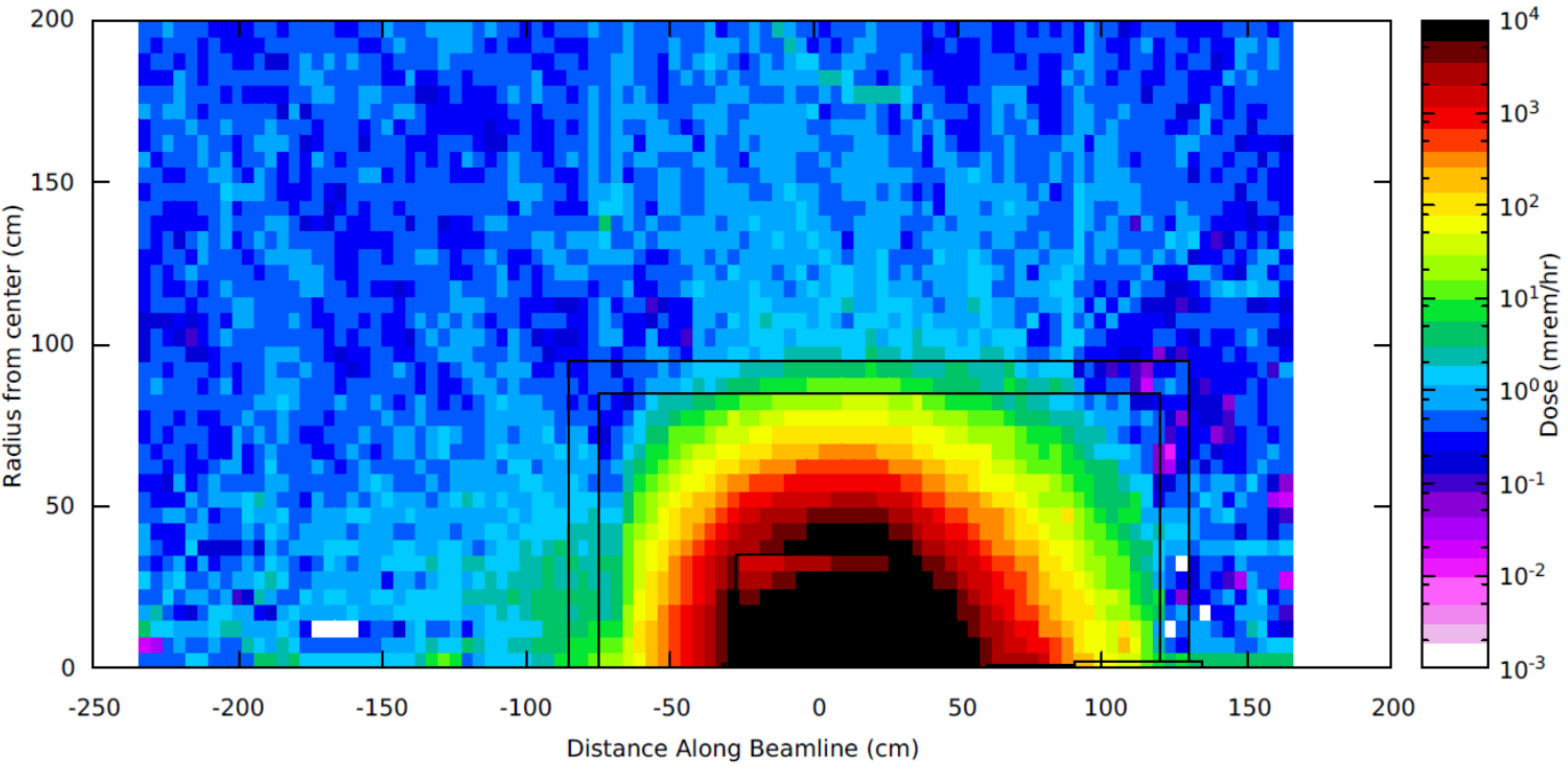}
}
\hspace{0.25cm}
\subfigure[\label{fig:activation-cps-radial} \it ] 
%caption for subfigure b
{
\includegraphics[width=2.8in]{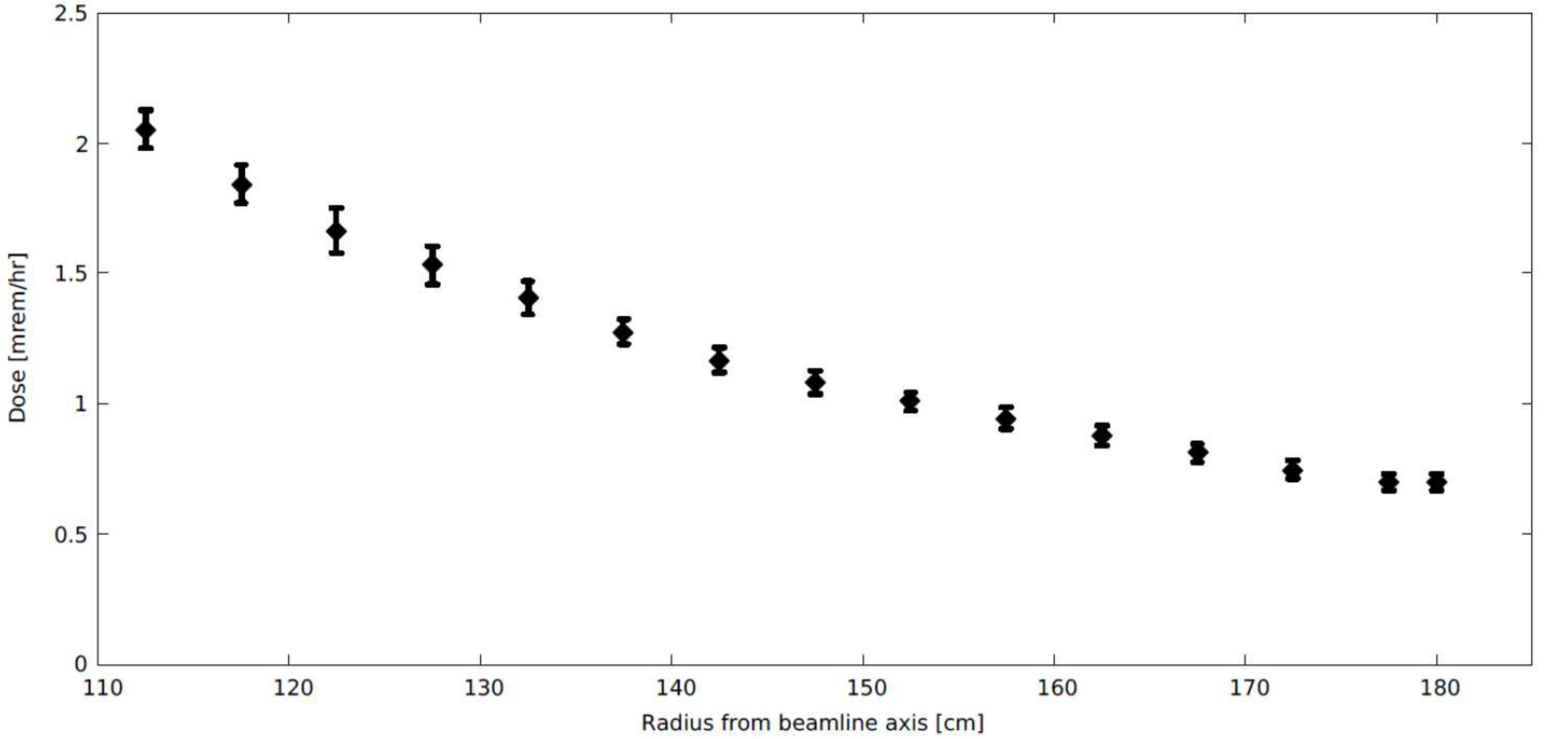}
}
\caption{\label{fig:activation-figures} Activation radiation dose rate
  one hour after a 1000-hour experiment as a function of position in
  the experimental hall for the case of a 2.7~$\mu$A electron beam
  hitting a 10\% copper radiator inside the CPS.}
\end{figure}

Dose rates due to the decay of activation products produced in the CPS
during beam-on conditions have been calculated.
Fig.~\ref{fig:activation-cps} shows the calculated activation dose one
hour after a 1000-hour experiment has been completed with the same
conditions as before (2.7 $\mu$A, 10\% copper radiator, with shielded
CPS). Fig.~\ref{fig:activation-cps-radial} shows the activation dose
rate as a function of radial distance from the CPS. The activation
dose outside the CPS is 2~mrem/hr at the surface and reduces radially
outward.  At a distance of one foot it is reduced to about
1.5~mrem/hr.  This therefore demonstrates that the current design
meets the requirement that the activation dose outside the device
envelope at one foot distance is $\le$ several mrem/hr after one hour
following the end of a 1000 hour run.

Note that these estimates do not depend much on the assumed 1000-hour
continuous running assumption, as similar dose rates are seen in a
calculation for a 100-hour continuous run, reflecting the fact that
much of the activation products are relatively short-lived.
Furthermore, activation dose rates do not drop appreciably after one
hour or even one day. On the other hand, after one month the
activation dose rates at the CPS surface are reduced by up to a factor
of ten. Inside the CPS the activation dose rate can be up to
1~krem/hr, which is why the CPS will be moved laterally to the side
after an experiment rather than disassembled.

        \subsection{Comparison with Dose Rates from the Target}
        
\begin{figure}
\centering
\includegraphics[width=3.0in]{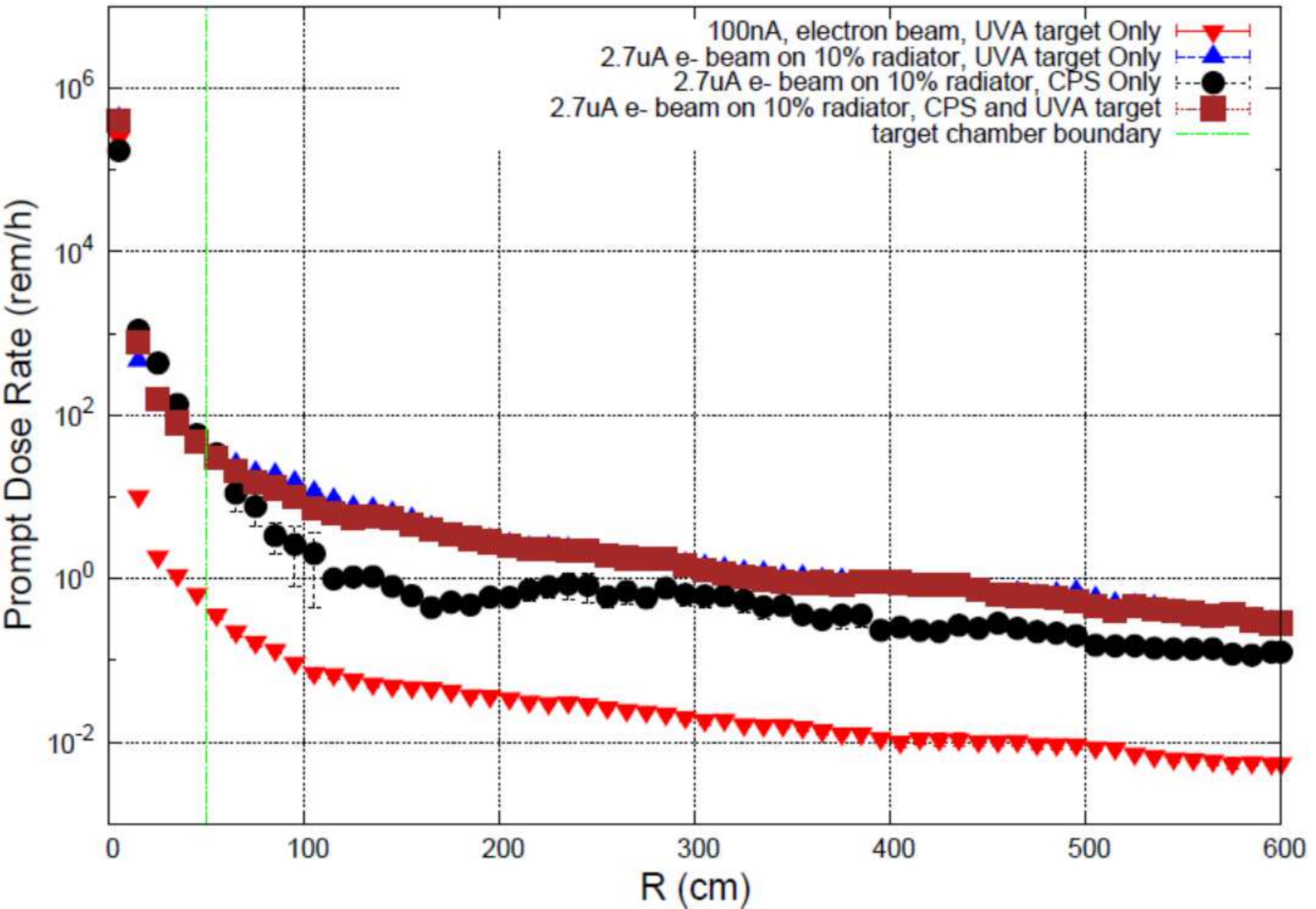}
\caption{\label{fig:prompt-at-target} \it Prompt dose at the
          target for different configurations. Distance R is radial
          distance from the target centre, with the radius of the
          scattering chamber boundary located at 50 cm.}
\end{figure}

        Fig.~\ref{fig:prompt-at-target} shows the prompt dose at the
        target for different experimental configurations as a function
        of radial distance from the target centre. It is worth
        commenting on the results for three of these configurations:
        the {\it 100 nA electron beam}, the {\it 2.7 $\mu$A photon
          beam} and the {\it CPS with polarized target}. At the
        boundary of the scattering chamber in the {\it 100 nA electron
          beam} configuration, the default operating mode for
        polarized beam experiments with dynamically nuclear polarized
        targets at Jefferson Lab to date, the prompt dose at the
        target is roughly 1~rem/hr. In the {\it 2.7 $\mu$A photon
          beam} scenario it is roughly 30~rem/hr, which simply
        reflects the fact that even if a 2.7 $\mu$A pure photon beam
        deposits the same heat load in a target as a 100 nA electron
        beam, the radiation rate is much higher. The {\it CPS with
          polarized target} scenario is identical to the pure photon
        beam case, further demonstrating that no additional radiation
        in the target area is created due to the presence of the CPS.

\begin{figure*}
\centering
\includegraphics[width=5.0in]{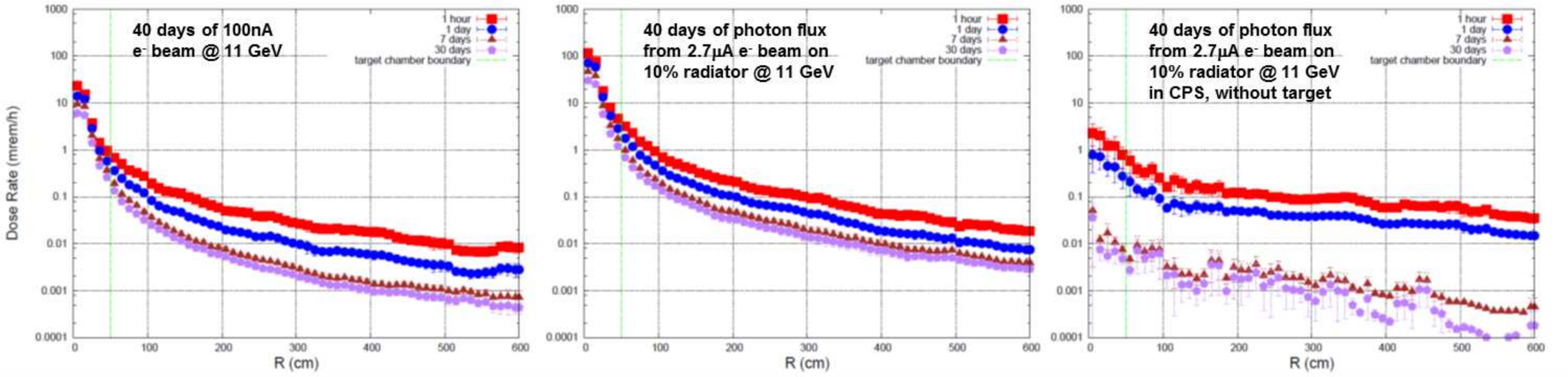}
\caption{\label{fig:activation-at-target} \it Activation dose rate
          at the target for different configurations. Distance $R$ is
          radial distance from the target centre, with the radius of
          the scattering chamber boundary located at 50 cm.}
\end{figure*}

        Similarly, Fig.~\ref{fig:activation-at-target} shows the
        activation dose rates for the same three configurations. One
        can see that the {\it 2.7 $\mu$A photon beam} configuration
        has a much higher activation dose rate at the target than the
        {\it 100 nA electron beam} case. This again reflects what was
        seen in the previous figure for the prompt radiation dose
        rate, as there are many more photons coming from a 2.7~$\mu$A
        electron beam on a 10\% copper radiator than there are from a
        100~nA electron beam on a roughly 3\% dynamically nuclear
        polarized target.  The effect of the CPS on the activation
        rate at the target is, as before, negligible.

\begin{figure}
\centering
\includegraphics[width=3.0in]{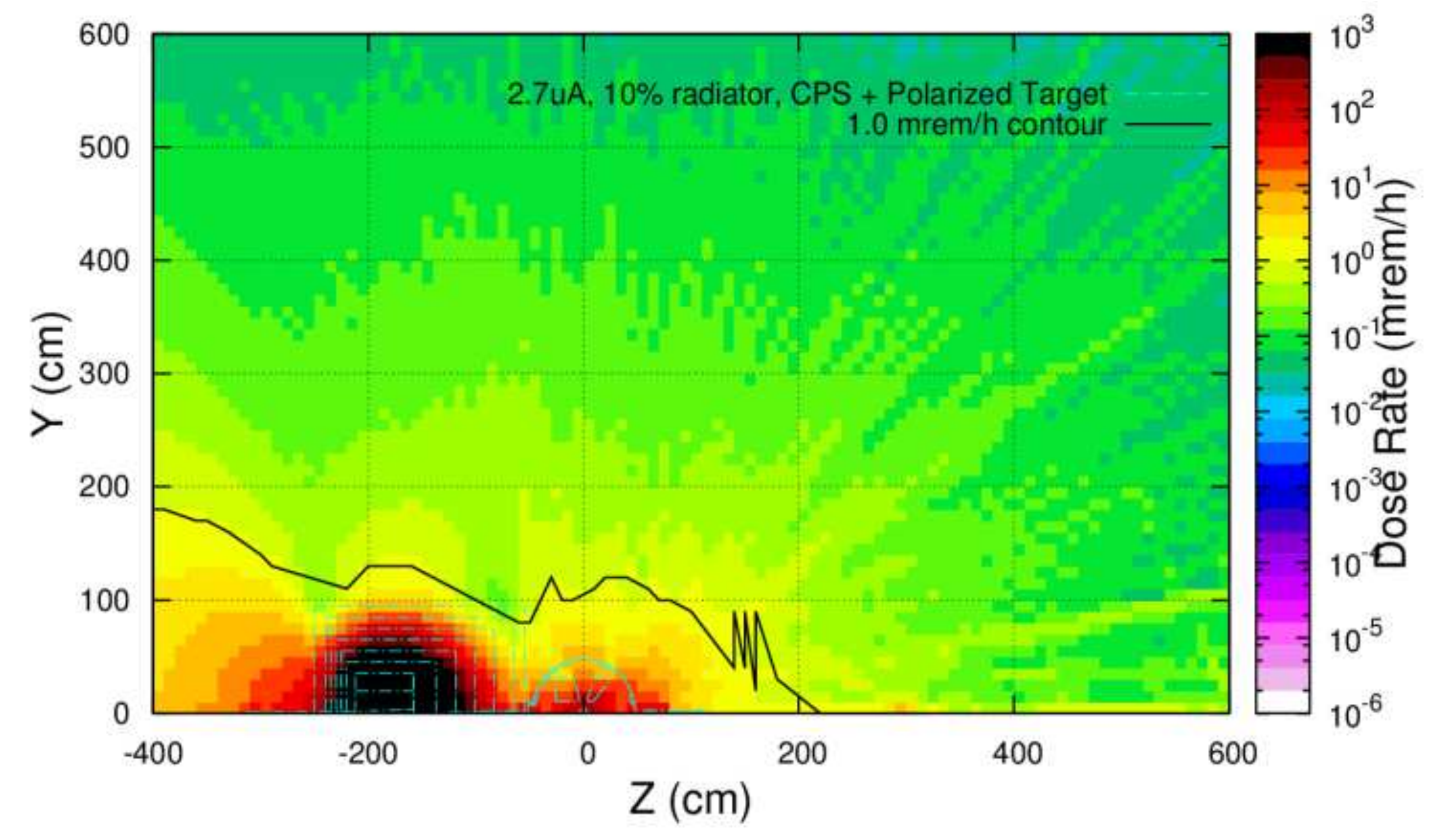}
\caption{\label{fig:activation-rates} \it Activation radiation
          dose rate one hour after a 1000-hour experiment as a
          function of position in the experimental hall for the case
          of a 2.7~$\mu$A electron beam hitting a 10\% copper radiator
          inside the CPS, with the target geometry included.  The
          1~mrem/hr contour is indicated.}
\end{figure}

        Fig.~\ref{fig:activation-rates} shows a two-dimensional plot
        of the activation dose rate in the experimental hall one hour
        after a 1000 hour run with the CPS, a 2.7~$\mu$A, 11~GeV beam
        on a 10\% radiator and the polarized target system (at z = 0).
        The 1~mrem/hour contour is indicated, and demonstrates that
        with the current CPS baseline design, the activation dose at
        the target centre in the experimental target area, where operational
        maintenance tasks may be required, is dominated by the dose
        induced by a pure photon beam.  At a distance of one foot from
        the scattering chamber it is $\le$ several mrem/hr one hour
        after a 1000 hour run, as required.  

        \subsection{Material Considerations}

\begin{figure*}
\centering
\includegraphics[width=5.0in]{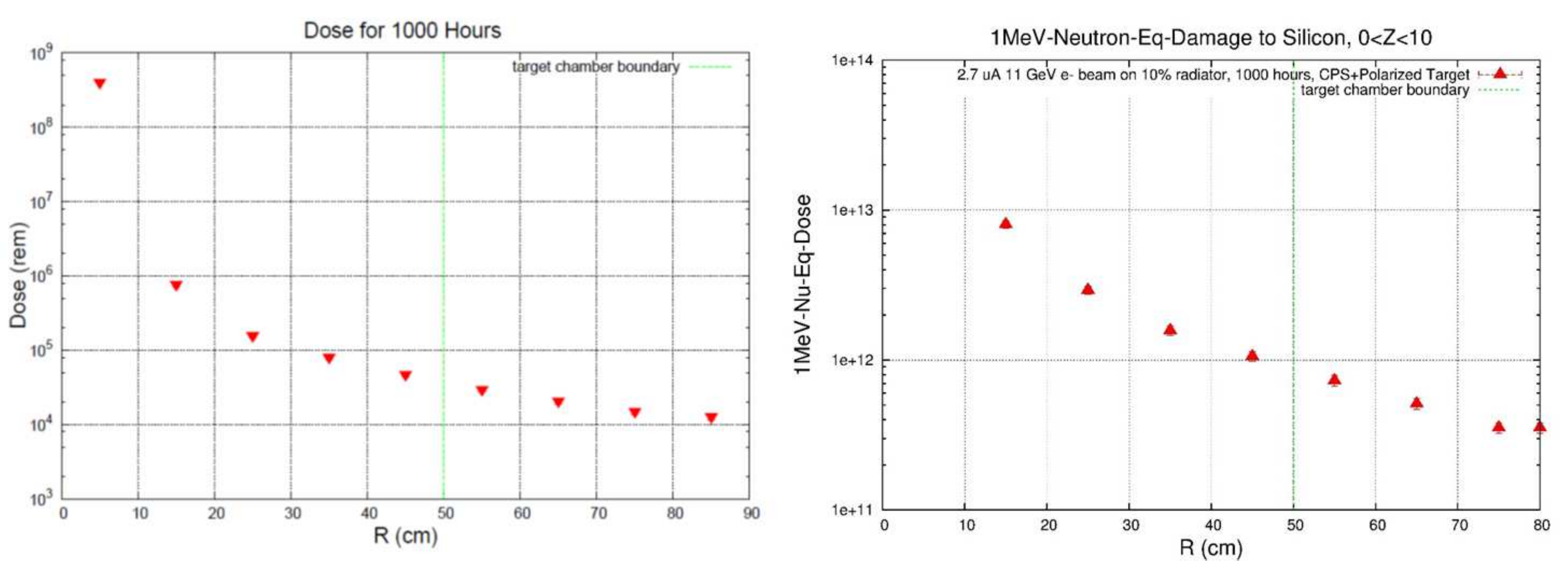}
\caption{\label{fig:prompt-and-neutron-1meveq-damage} \it The prompt radiation dose (left) and the resulting
          1~MeV neutron equivalent damage to silicon (right) in the
          target area, assuming the conditions described above.  The
          polarized target system is centred at $R = 0$, the nominal
          target chamber radius is 50~cm and the target coils are at
          about 20~cm from the beam line. The dose at the target coils
          is $5\times 10^5$~rem and the 1~MeV neutron equivalent
          damage is $5 \times 10^{12}$~neutrons/cm$^2$.}
\end{figure*}

        The level of radiation of the CPS experiments is well below
        what is typical for many high-luminosity experiments at
        Jefferson Lab using regular cryogenic target systems and/or
        radiators. However, the radiation level on the polarized
        target coils, due to the interaction of the photon beam with
        the polarized target material, will be higher than in previous
        experiments (around 500~rem/hr as illustrated in
        Fig.~\ref{fig:prompt-and-neutron-1meveq-damage}). This is not
        expected to pose any significant issues. Furthermore, the
        radiation levels in the CPS magnet coils at a distance of
        20~cm from the radiation source are around 1~Mrem/hr (see {\sl
          e.g.} Fig.~\ref{fig:prompt_cps-1D}, bottom right). This
        relatively moderate level will allow the use of a modest-cost
        Kapton tape-based insulation of the coils~\cite{Kapton}.

\section{Engineering and Safety Aspects}
\label{sec:safety-and-engineering}

As stated earlier, cooling of the CPS core will require four gallons
of water per minute at 110 psi pressure, which will result in a
30$^\circ$C rise in coolant temperature. Activation of this coolant
water and beam dump is anticipated, meaning a closed-cycle cooling
system will be needed.  Activation inside the CPS will be confined to
a very small volume and in the event of a leak, external contamination
will be minimized. A leak pan under the device could easily be
included to catch and confine any leakage up to and including a total
loss of primary coolant. A modular pallet mounted design would be
efficient and would include primary coolant pumps, DI resin beds, heat
exchanger, surge tank, controls instrumentation and manifolds.

The combination of placing a high-power bremsstrahlung radiator, a
magnet and a beam dump inside a shielded box imposes significant
reliability and remote handling considerations. The primary
engineering control involves making the design as robust as possible,
including large safety margins and avoiding the need for disassembly
for maintenance or any other reason.  The CPS should be heavily
instrumented for early detection of problems such as low coolant flow,
leaks, low pressure, high temperature, and high conductivity. The two
areas where conservative safety design is most needed are in the
magnetic coil and dump cooling systems.

A low magnet coil current density design is envisioned, which is not
expected to exceed 500 A/cm$^2$. In order to allow easy access,
individual coil pancake leads should be extended to an area outside of
the magnet and shielding. There should be no electrical or coolant
joints inside the shielding, and each separate sub-coil of the CPS
magnet should have thermometers, thermal circuit breakers, voltage and
coolant flow monitors to avoid any possibility that one of the
separate current paths can overheat due to lack of sufficient coolant
or a bad electrical contact.  Extra insulation between sub-coils and
between the coil and ground should be added to prevent ground faults.
Lastly, a commercial power supply is assumed that will come with a
wide array of internal interlock protections. The available interlocks
and signals can be fed into the electron beam Fast Shutdown (FSD)
system.

To protect equipment in the experimental hall from the beam striking
the CPS shielding, a dual protection scheme using both a beam position
monitoring system and direct instrumentation of the fast raster magnet
is proposed. The beam diagnostics systems would monitor beam position
and motion in close to real time and monitor coild voltage on the
raster coils, which would provide ample early warning of raster
problems.  Both of these independent signals would be fed into the FSD
system.  Radiator temperature could be monitored to provide a third
independent protection system, and if implemented, thermocouples
mounted on the radiator should be robust against radiation damage and
provide fast enough protection against radiator overheating.

\section{Summary}
\label{sec:summary}

The Compact Photon Source (CPS) design features a magnet, a central
copper absorber and hermetic shielding consisting of tungsten powder
and borated plastic.  The addition of the latter has a considerable
impact on reducing the neutron flux escaping the CPS. The ultimate
goal in this design process is that radiation from the source should
be a few times less than from a photon beam interacting with the
material of a polarized target. The equivalent heat load for a pure
photon beam impinging such targets corresponds to a photon flux
originating from a 2.7 $\mu$A electron beam current striking a 10\%
copper radiator. Detailed simulations of the power density and heat
flow analysis show that the maximum temperature in the absorber is
below 400 degrees, which is well within the acceptable range of
copper, and thus demonstrates that the CPS can absorb 30 kW in total,
{\sl e.g.} corresponding to an 11-GeV electron beam energy and a 2.7
$\mu$A electron beam current.

The CPS also fulfills the requirements on operational dose rates at
Jefferson Lab, which have been established with extensive and
realistic simulations. The projected prompt dose rate at the site
boundary is less than 1 $\mu$rem/hr (to be compared with 2.4
$\mu$rem/hr, which corresponds to a typical JLab experiment that does
not require extra shielding). The activation dose outside the device
envelope at one foot distance is less than several mrem/hr after one
hour following the end of a 1000 hour run ($\sim$ 3 months). The
activation dose at the target centre in the experimental target area, where
operational maintenance tasks may be required, is dominated by the
dose induced by the pure photon beam. At a distance of one foot from
the scattering chamber it is less than several mrem/hr one hour after
the end of a 1000 hour run (i.e. the additional activation dose
induced by absorption of the electron beam in the Compact Photon
Source is negligible).

\section{Acknowledgements}
This work is supported in part by the National Science Foundation grants PHY-1306227, 1913257, and PHY-1714133, the U.S. Department of Energy, Office of Science, Office of Nuclear Physics award DE-FG02-96ER40950, and the United Kingdom’s Science and Technology Facilities Council (STFC) from Grant No. ST/P004458/1.
We would like to thank Paul Brindza for helpful discussions and providing valuable input for the writing of this document.
This material is based upon work supported by the U.S. Department of Energy, Office of Science, Office of Nuclear Physics under contract DE-AC05-06OR23177.

\end{document}